\documentclass{IEEEtran}
\usepackage{siunitx}
\usepackage{commath}
\usepackage{graphicx}
\usepackage{textcomp}
\usepackage{cite}

\begin{document}
\bstctlcite{IEEEexample:BSTcontrol} 
\title{Simulation of Silicon Waveguide Single-Photon Avalanche Detectors for Integrated Quantum Photonics}

\author{\IEEEauthorblockN{Salih Yanikgonul,
Victor Leong,
Jun Rong Ong,
Ching Eng Png, and
Leonid Krivitsky}

\thanks{ 
This work was supported by NRF-CRP14-2014-04, “Engineering of a Scalable Photonics
Platform for Quantum Enabled Technologies”. 
S. Yanikgonul was supported by the Singapore International Graduate Award (SINGA). 
(Corresponding author: someone)

S. Yanikgonul, V. Leong, and L. Krivitsky are with the Institute of Materials Research and Engineering, Agency for Science, Technology and Research (A*STAR), 13834, Singapore.
S. Yanikgonul is also with the School of Electrical and Electronic Engineering, Nanyang Technological University, 639798, Singapore (e-mail: Salih\_Yanikgonul\_from.tp@imre.a-star.edu.sg; victor\_leong@imre.a-star.edu.sg; Leonid\_Krivitskiy@imre.a-star.edu.sg).

J. R. Ong and C. E. Png are with the Institute of High Performance Computing, Agency for Science, Technology and Research (A*STAR), 138632, Singapore. (e-mail: ongjr@ihpc.a-star.edu.sg; pngce@ihpc.a-star.edu.sg)
}}

\IEEEtitleabstractindextext{
\begin{abstract}
Integrated quantum photonics, which allows for the development and implementation of chip-scale devices, is recognized as a key enabling technology on the road towards scalable quantum networking schemes.
However, many state-of-the-art integrated quantum photonics demonstrations still require the coupling of light to external photodetectors. 
On-chip silicon single-photon avalanche diodes (SPADs) provide a viable solution as they can be 
seamlessly integrated with photonic components, and operated with high efficiencies and low dark counts at temperatures achievable with thermoelectric cooling. Moreover, they are useful in applications such as LIDAR and low-light imaging. 
In this paper, we report the design and simulation of silicon waveguide-based SPADs on a silicon-on-insulator platform for visible wavelengths, 
focusing on two device families with different doping configurations: \mbox{p-n$^{+}$} and \mbox{p-i-n$^{+}$}. 
We calculate the photon detection efficiency (PDE) and timing jitter 
at an input wavelength of 640\,nm 
by simulating the avalanche process using a 2D Monte Carlo method, 
as well as the dark count rate~(DCR) at 243\,K and 300\,K.
For our simulated parameters, 
the optimal \mbox{p-i-n$^{+}$} SPADs show the best device performance, 
with a saturated PDE of $\textbf{52.4}\pm\textbf{0.6}$\% at a reverse bias voltage of 31.5\,V, full-width-half-max (FWHM) timing jitter of~10\,ps, and a DCR of $<$\,5 counts per second at 243\,K. 

\end{abstract}

}

\maketitle
\IEEEdisplaynontitleabstractindextext

\section{Introduction}
Quantum information technologies have been rapidly developing in recent years, and efforts are shifting from conceptual laboratory demonstrations to scalable real-world devices~\cite{Wehner2018}.
Chip-scale photonics devices are important candidates for implementing key features of a future quantum internet,
but many recent demonstrations still require the coupling of light to external single-photon detectors~\cite{Qiang2018,Sibson2017}. 
Major improvements in device footprint and scalability could be achieved 
if these photodetectors reside on the same chip and couple directly to the photonic waveguides~\cite{OBrien2009}.

Superconducting nanowire single-photon detectors (SNSPDs) are a state-of-the-art solution, 
featuring waveguide integrability, near-unity quantum efficiencies, low dark count rate of a few counts per second (cps), and low timing jitter down to $<$\,20\,ps~\cite{najafi2015chip,Akhlaghi2015}.
However, they require cryogenic operating temperatures of a few degrees Kelvin, which is expensive and prohibitive for large-scale deployment.

A practical alternative can be found in single-photon avalanche diodes (SPADs), 
which are typically reverse biased beyond the breakdown voltage.
In this so-called Geiger mode, a single incident photon can trigger a macroscopic avalanche current
via a cascade of impact ionization processes.
In contrast to SNSPDs, SPADs typically only require thermoelectric cooling and can even operate at room temperature~\cite{warburton2009free,liang2017room}.
Moreover, SPADs can be easily incorporated into silicon photonics platforms and benefit from mature complementary metal-oxide semiconductor (CMOS) fabrication technologies~\cite{sacher2015multilayer}, making them a promising candidate for scalable manufacturing.

To date, reports of waveguide-coupled SPADs have been limited to operation at infrared wavelengths~\cite{Martinez:16,Martinez:17}.
However, many relevant quantum systems, including trapped ions~\cite{mehta2016integrated} and color centers in diamond~\cite{sipahigil2016integrated}, operate in the visible spectrum, 
which makes efficient, low-noise SPADs for visible wavelengths highly desirable.
Such devices would also find numerous applications in other important technologies, including
LIDAR~\cite{Du2018},
non-line-of-sight imaging~\cite{OToole2018},  
and fluorescence medical imaging~\cite{Homulle2016}. 

In this paper, we extend our recent work on the design and simulation of silicon waveguide-coupled SPADs for visible light operation, where we used a 2D Monte Carlo simulator to obtain the photon detection efficiency (PDE) and timing jitter, and studied the effect of different waveguide dimensions and doping concentrations~\cite{Yanikgonul2018}. 
Here we perform an in-depth study of different doping configurations, focusing on two device families: \mbox{p-n$^{+}$} and \mbox{p-i-n$^{+}$}.
In addition to the PDE and timing jitter, we also analyze the expected dark count rate (DCR) at room temperature and at -30\,\si{\celsius} (243\,K), which is a typical operating temperature achievable by Peltier coolers. 

Many details regarding the basic SPAD geometry and simulation procedure can be found in ref.~\cite{Yanikgonul2018} and are not repeated here; instead we provide the essential points and highlight the improvements we have made on our previous work.

\section{Waveguide-coupled SPAD Designs}
\subsection{Device Geometry}
\begin{figure}[tb]
    \centering
    \includegraphics[width=1\linewidth]{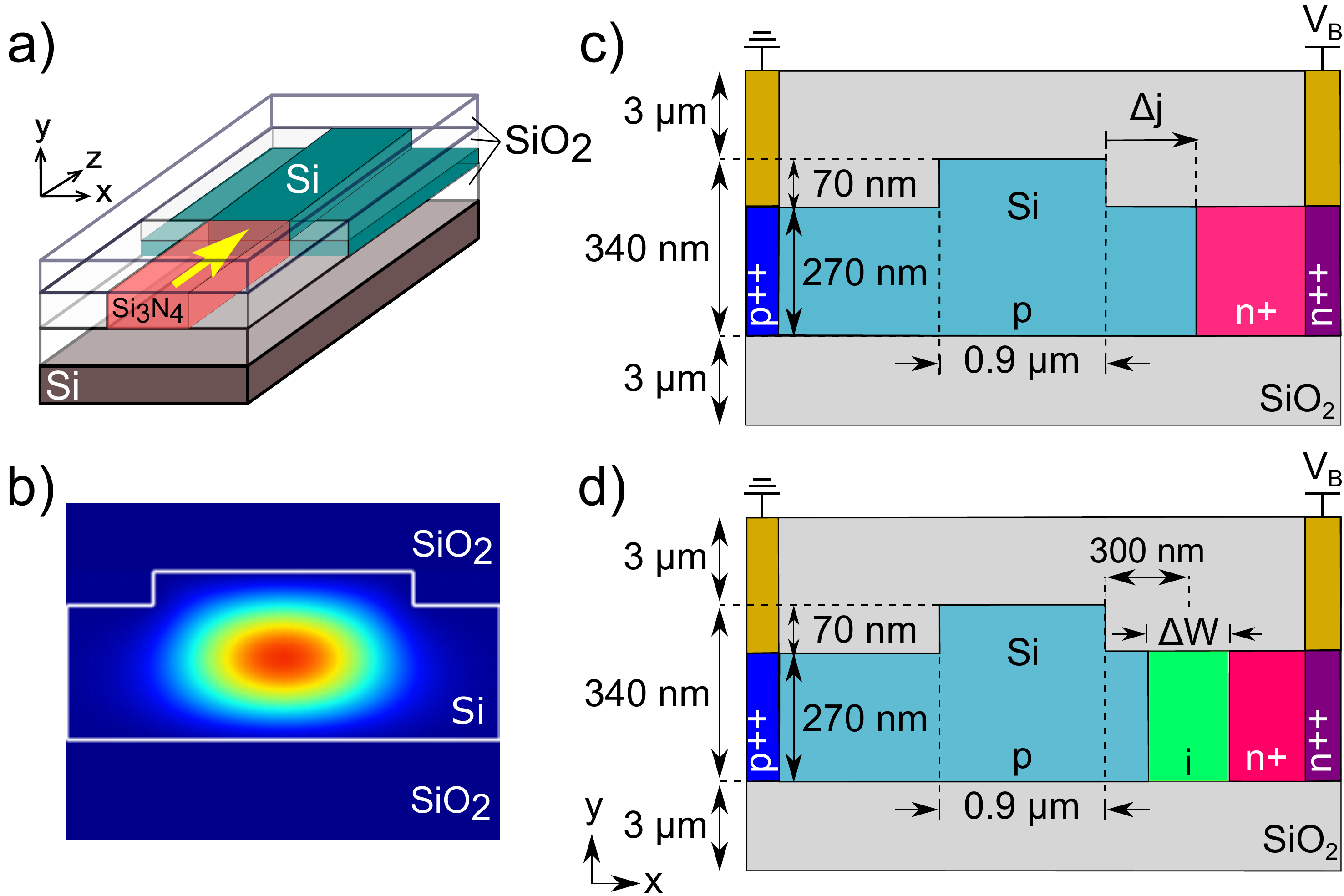}\\
    \caption{(a)~SPAD structure, consisting of a Si rib waveguide end-fire coupled to an input Si$_{3}$N$_{4}$ waveguide,
             (b)~optical mode profile at 640 nm for the fundamental (quasi-)TE mode, 
             (c)~\mbox{p-n$^{+}$} doping configuration with the junction placed at a distance $\Delta j$ from the right edge of the waveguide core, 
             (d)~\mbox{p-i-n$^{+}$} doping configuration with an intrinsic region width $\Delta W$. 
             The cross section is constant along the length of the waveguide. Images are not drawn to scale.}
    \label{fig:SPAD_design}
\end{figure}

The SPAD structure is shown in Fig.~\ref{fig:SPAD_design}. 
It is based on a silicon-on-insulator (SOI) platform, and consists of a 16\,\si{\micro\metre} long silicon (Si) rib waveguide with 
an absorption of $>$99\% at 640 nm.
Input light is end-fire coupled from an input silicon nitride (Si$_3$N$_4$) rectangular waveguide, which has high transmittivity at visible wavelengths~\cite{sacher2015multilayer}. 
We choose this input coupling geometry over a phase-matched interlayer transition, commonly used in integrated photodetectors for infrared wavelengths~\cite{Martinez:17,Chen:17}, as the latter is difficult to achieve due to the large difference in refractive indices for Si ($n$\,=\,3.8) and Si$_3$N$_4$ ($n$\,=\,2.1). 
An input coupling efficiency of $>$90\% at the Si/Si$_3$N$_4$ interface is obtained using 3D Finite Difference Time Domain (FDTD) simulations (Lumerical). 

The structures are cladded with 3\,\si{\micro\metre} of silicon dioxide (SiO$_2$) above and below. 
In this study, we fixed the waveguide core width and height at 900 nm and 340 nm respectively, with a shallow etch giving a rib height of 270 nm.

Electrical connections to the device would be made via metal electrodes deposited on top of heavily-doped p$^{++}$ and n$^{++}$ regions 
at the far ends of the device (along the $x$ axis).

\subsection{Doping Configurations}
\label{sec:doping_configuration}
Our previous simulation study of \mbox{p-n$^+$} SPADs~\cite{Yanikgonul2018} showed that increasing waveguide core widths (up to 900\,nm) could lead to a higher PDE,
as charge carriers can travel a larger distance over which avalanche multiplication can occur.
Here, 
we vary the placement of the p-n$^+$ junction,
and investigate the hypothesis that increasing the displacement~$\Delta j$ of the junction beyond the edge of the waveguide core region (Fig.~\ref{fig:SPAD_design}(c)) would also enhance this effective distance, and hence the PDE.

Another observation was that impact ionization was most efficient in a narrow region where the highest electric fields are concentrated~(similar to Figs.~\ref{fig:Electric_fields}(a)-(c)).
Widening this high-field region could enhance the PDE, and is achievable by introducing an intrinsic region between the p- and \mbox{n$^+$-doped} areas~(Fig.~\ref{fig:SPAD_design}(d)).
However, doing so would also lower the peak electric field strength~(Fig.~\ref{fig:Electric_fields}(d)), which could in turn decrease the impact ionization efficiency.
Here we explore the effectiveness of such \mbox{p-i-n$^+$} devices, and attempt to find the optimum width of the intrinsic region~$\Delta W$, centered at 300\,nm from the edge of the waveguide core.

For both device families, we maintain a constant geometry and doping profile along the length of the waveguide.
In this study, we choose a n$^+$ (p) doping concentration of 1$\times$10$^\textrm{19}$ (2$\times$10$^\textrm{17}$) dopants/cm$^\textrm3$, and a lightly \mbox{p-doped} intrinsic region with 1$\times$10$^\textrm{15}$ dopants/cm$^\textrm3$.

\section{Simulation Method}

\subsection{DC Electrical Analysis}

For each set of device dimensions and doping configurations, 
we perform a DC electrical analysis (ATLAS, Silvaco Inc.) by applying a reverse bias voltage $V_B$ across the device electrodes.
For each device, the cathode and anode are placed  
equidistant from the center of the Si waveguide, 
with a minimum n$^+$ region width of 45\,nm.
We thus obtain the electric field $\mathbf{F}(\mathbf{r})$,
ionization coefficients, 
and other parameters dependent on the 2D position vector~$\mathbf{r}$ in the $x-y$ plane; these are required for the Monte Carlo simulation of the avalanche process.
Further details can be found in ref.~\cite{Yanikgonul2018}.
The breakdown voltage is also identified as the reverse bias voltage $V_B$ at which the device current increases sharply.

\begin{figure}[tb]
    \centering
    \includegraphics[width=0.64\linewidth]{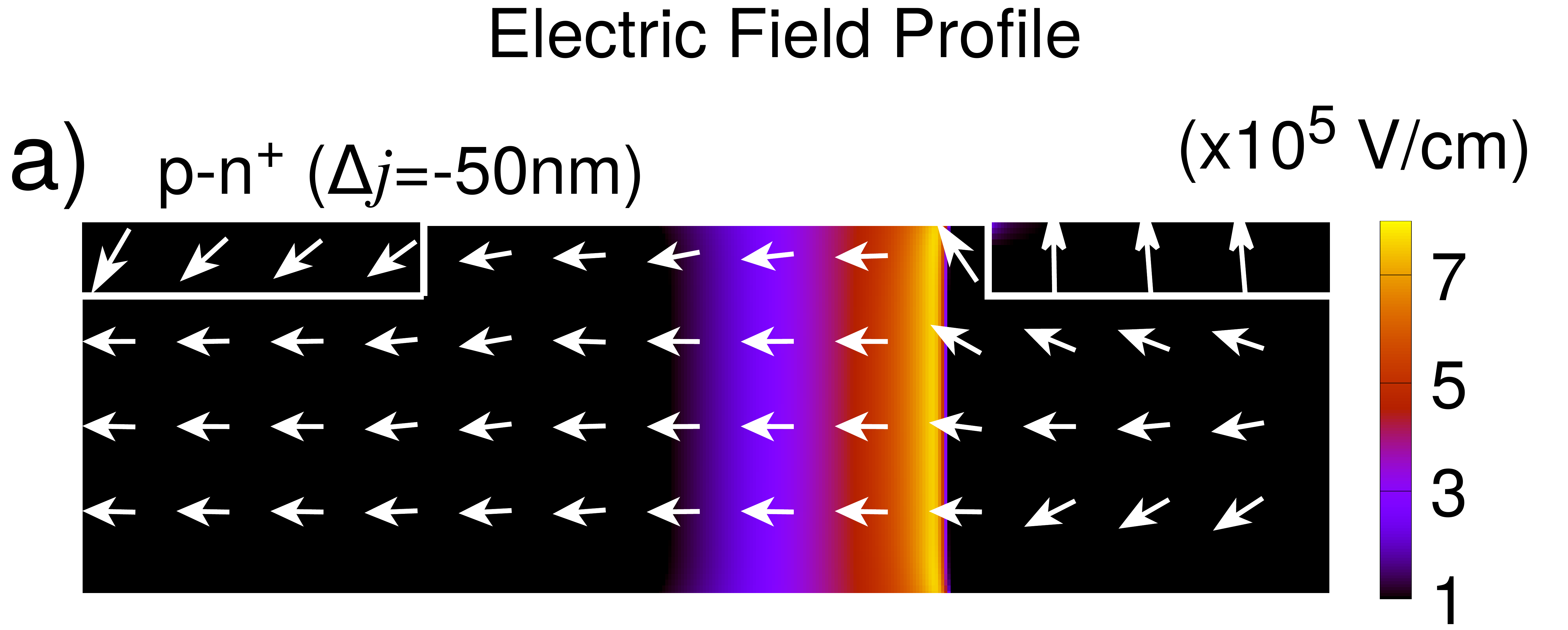} \includegraphics[width=0.34\linewidth]{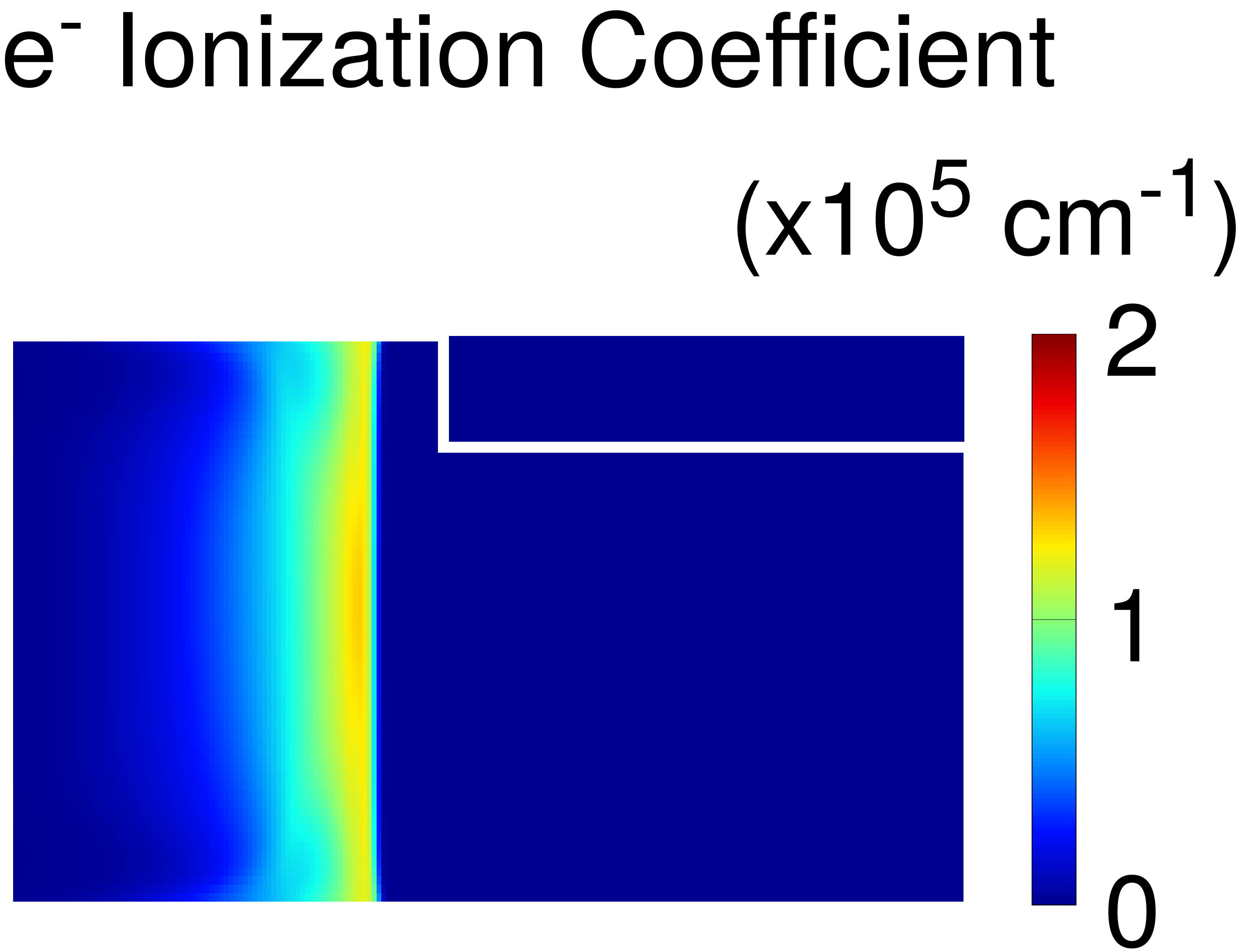}\vspace*{0.15cm}
    \includegraphics[width=0.64\linewidth]{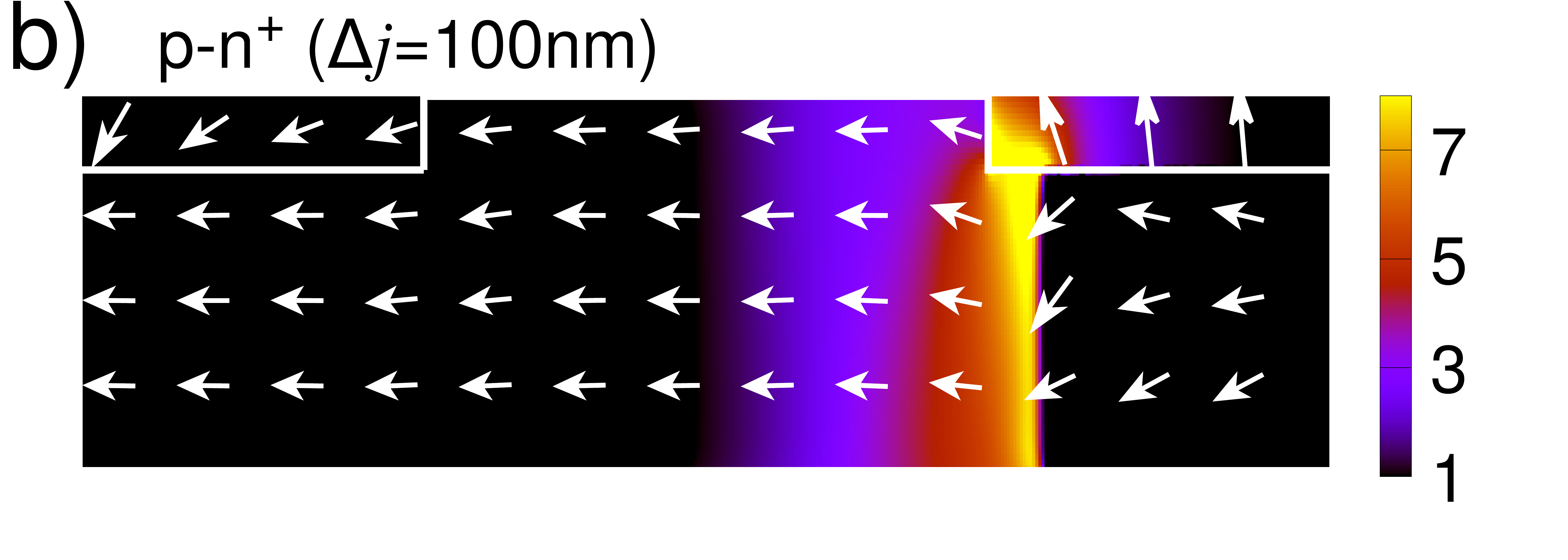} \includegraphics[width=0.34\linewidth]{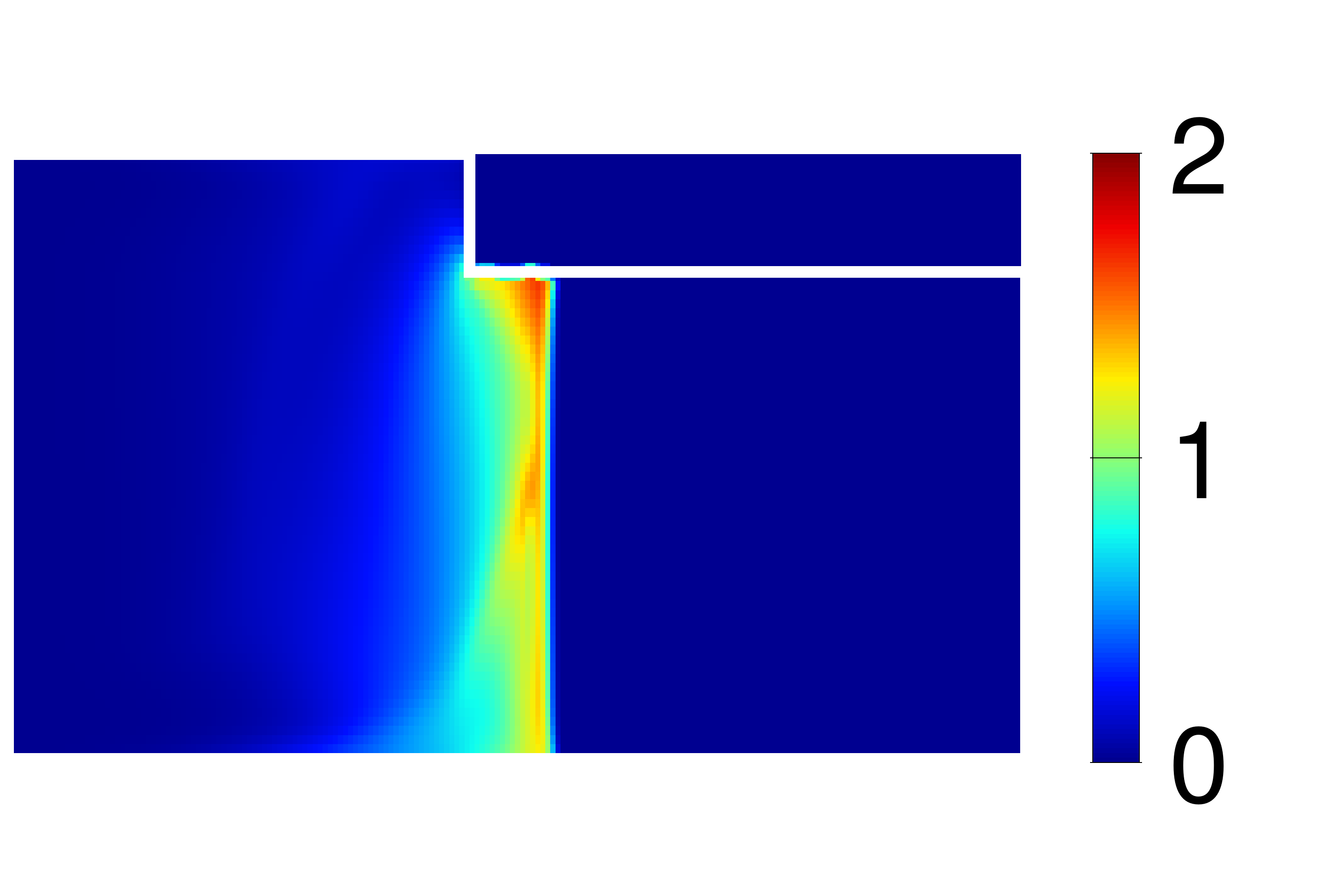}
    \includegraphics[width=0.64\linewidth]{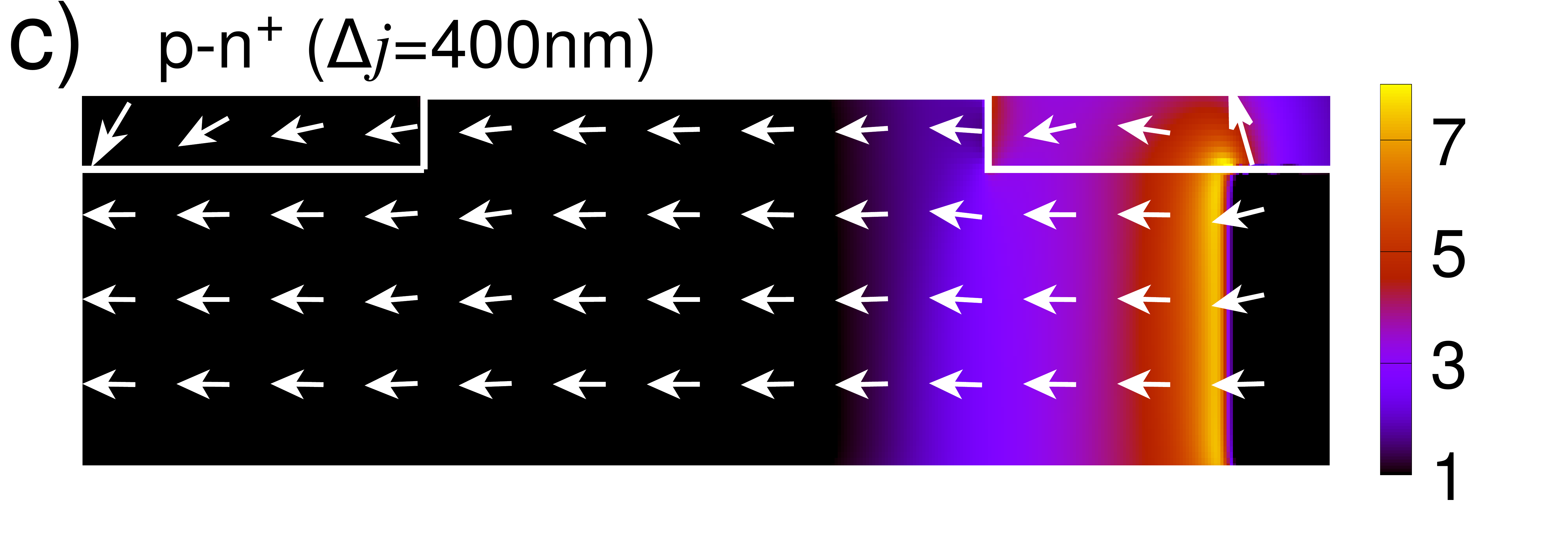} \includegraphics[width=0.34\linewidth]{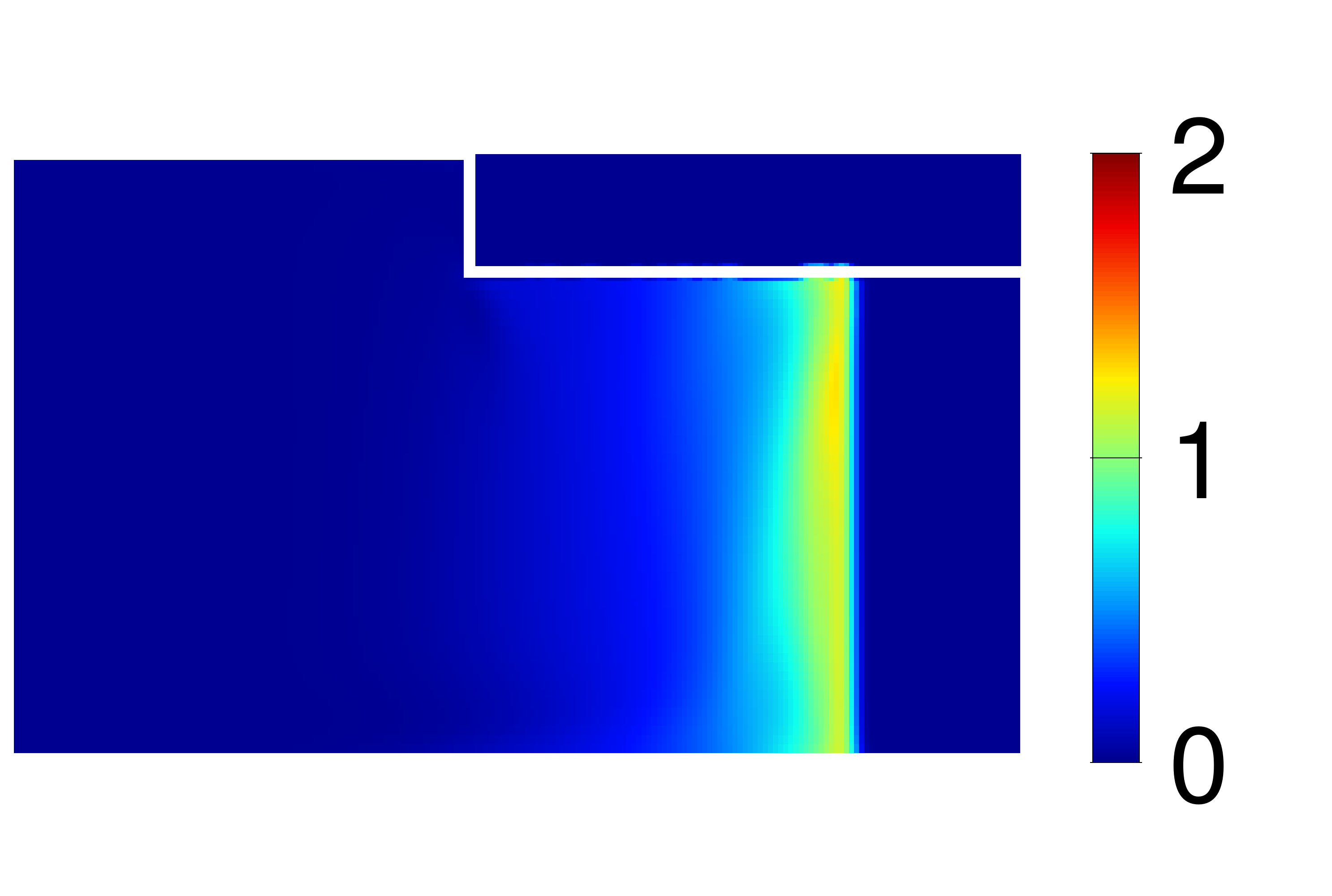}
    \includegraphics[width=0.64\linewidth]{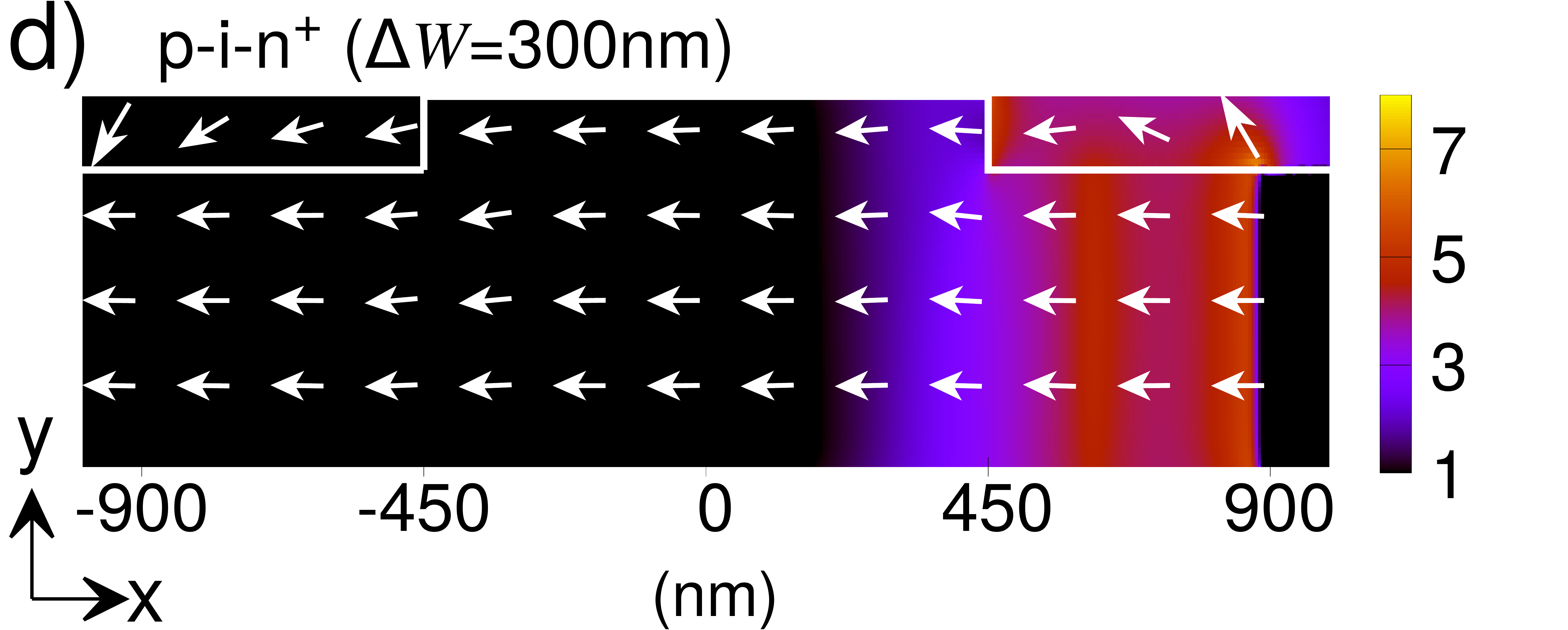} \includegraphics[width=0.34\linewidth]{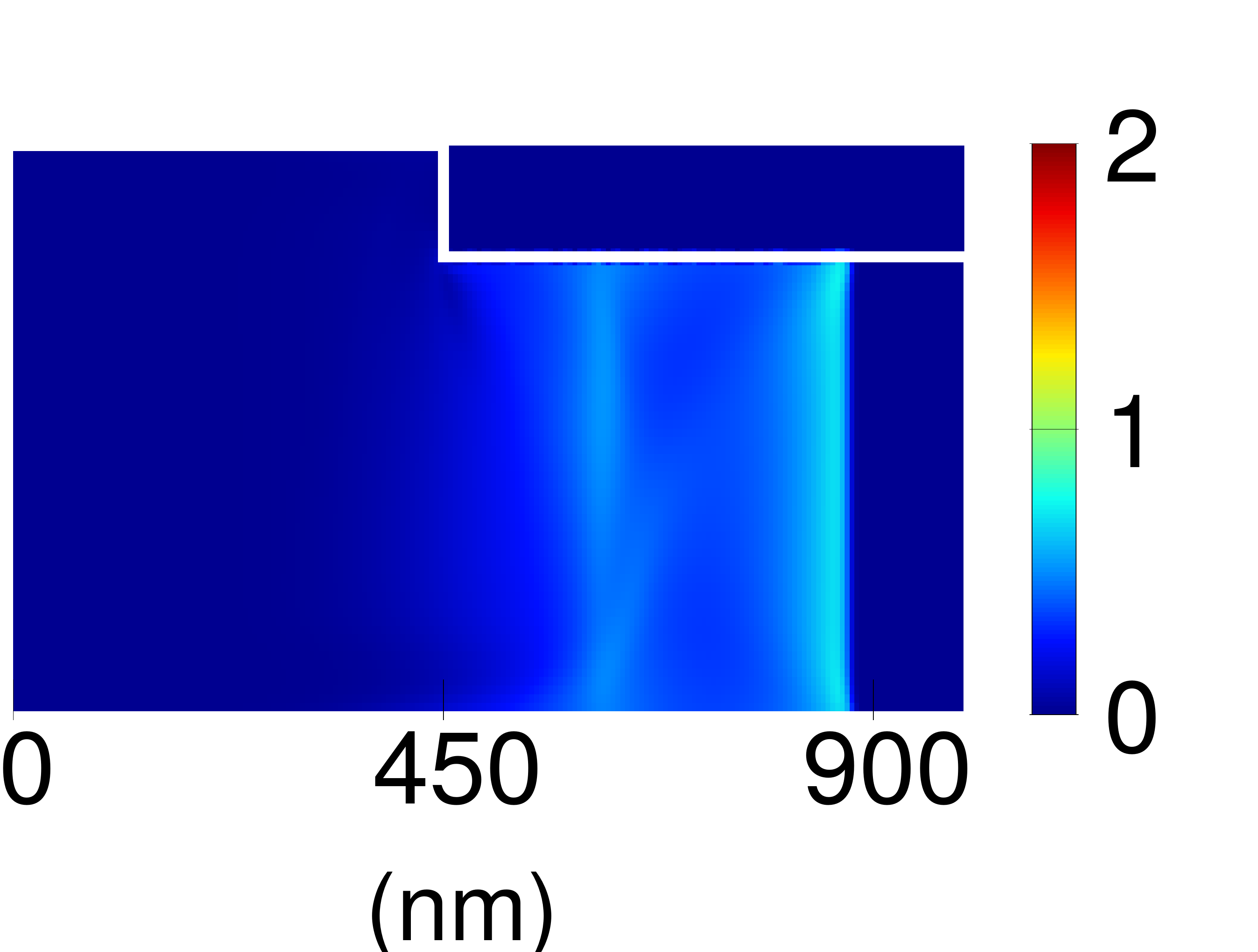}
    \caption{Electric field profiles (left) and electron ionization coefficients (right) at reverse bias voltages $V_B$ where the photon detection efficiency (PDE) saturates.
             a) \mbox{p-n$^{+}$} SPAD with $\Delta j$\,=\,-50\,nm at \mbox{$V_B$\,=\,18.5\,V},
             b) \mbox{p-n$^{+}$} SPAD with $\Delta j$\,=\,100\,nm at \mbox{$V_B$\,=\,23.7\,V},
             c) \mbox{p-n$^{+}$} SPAD with $\Delta j$\,=\,400\,nm at \mbox{$V_B$\,=\,26.5\,V}, and 
             d) \mbox{p-i-n$^{+}$} SPAD with $\Delta W$\,=\,300\,nm at \mbox{$V_B$\,=\,28.2\,V}. 
             For the electron ionization coefficients, only half the waveguide ($x$\,$>$\,0) is shown as their values are negligible in the other half.
             }
    \label{fig:Electric_fields}
\end{figure}
\subsection{2D Monte Carlo Simulator}

In comparison to deterministic techniques~\cite{Gulinatti2009}, Monte Carlo simulators are well-suited for analyzing SPAD performance, as they can evaluate the timing jitter by modeling the stochastic nature of the impact ionization and avalanche buildup processes.
For applications such as quantum key distribution (QKD)~\cite{collins2007low} and LIDAR~\cite{buller2007ranging}, low timing jitter is critical to the overall system performance.

In this work, we adapt the 2D Monte Carlo simulator detailed in ref.~\cite{Yanikgonul2018}.
Briefly, a random path length (RPL) model is used to simulate the avalanche multiplication process~\cite{Ingargiola2009,Ma2015,Ong1998a}.
Each simulation run starts with a photon absorption which creates an electron-hole pair. 
At each time step of interval~$\Delta t_{\textrm{rpl}}$,
each charge carrier is accelerated by the electric field and,
depending on the ionization coefficients, 
probabilistically causes an impact ionization after traversing a random path length.
This creates further electron-hole pairs, which can then undergo further impact ionizations and eventually lead to a self-sustaining avalanche.
Charge carriers are lost when they exit the device boundaries; 
we note that unlike in ref.~\cite{Yanikgonul2018}, the Monte Carlo simulation in this work considers the entire device area (the whole of the p, i, and n$^+$ regions) and is not restricted to the waveguide core region.

A successful detection event results if the device current reaches a detection threshold $I_\textrm{det}$.
Treating the success and failure outcomes as a binomial distribution, the PDE is then the fraction of successful detection events over all simulation runs, with an uncertainty given by the standard deviation~(s.d.). 
The distribution of avalanche times (i.e. time between photon absorption and reaching $I_\textrm{det}$) yields the timing jitter.

\subsubsection{Diffusion in Quasi-Neutral Regions}

The SPAD can be divided into a depletion region and quasi-neutral regions depending on the electric field strength.
In the depletion region, the dominant charge carrier transport process is the drift force due to the strong electric fields, and the RPL model applies.
However, in the quasi-neutral regions where electric fields are weak, impact ionization can be neglected, and a diffusion model which combines random walks (driven by Brownian motion) and the electric drift force is more suitable.
Similar to ref.~\cite{Yanikgonul2018},
we use a threshold field to define the quasi-neutral region, 
\mbox{i.e. $|\mathbf{F}(\mathbf{r})|<F_\textrm{thr}$\,=\,1$\times$10$^\textrm{5}$~V/cm,}
which is on the same order as the breakdown field in silicon~\cite{Wegrzecka}.

We use the fundamental (quasi-)TE mode profile~(Fig.~\ref{fig:SPAD_design}(b)) as a probability density map to determine the location where the initial electron-hole pair is injected for each simulation run.
If the injection occurs in the quasi-neutral regions, charge carrier transport is simulated using the diffusion model;
if the charge carrier crosses over to the depletion region, the simulation continues under the RPL model.

\subsubsection{Device Current via Shockley-Ramo's Theorem}

Ref.~\cite{Yanikgonul2018} calculates the device current using a 1D approximation of Ramo's theorem, 
which only considers the motion of charge carriers in one direction.
However, this would not be suitable here given our SPAD designs and more complex electric field profiles.
As such, we use the generalized Shockley-Ramo's current theorem~\cite{Shockley1938,Ramo1939}, 
where each charge carrier~$i$ at position~$\mathbf{r}_i$ contributes to the device current~$I$ induced on the cathode via:

\begin{equation}
I = \sum_{i}^{} q_i \cdot \mathbf{v}_{i}(\mathbf{r}_{i}) \cdot \mathbf{F}_0(\mathbf{r}_{i})
\label{eq:Induced_current}
\end{equation}
where $q_i$ is the charge, $\mathbf{v}_{i}(\mathbf{r}_{i})$ is the instantaneous velocity, and $\mathbf{F}_0(\mathbf{r}_{i})$ is a weighting electric field calculated 
in a similar way to $\mathbf{F}(\mathbf{r})$, but 
under these modified conditions:
(i) the cathode is at unit potential, while the anode is grounded; 
(ii) all charges (including space charges) are removed, i.e. the waveguide is undoped~\cite{He2001}.

\subsection{Dark Count Rate}

Even in the absence of light, free charge carriers may be generated, which can probabilistically trigger avalanche events and result in dark counts.
Due to the high electric fields in SPADs, 
the most relevant carrier generation mechanisms are
thermal excitation enhanced by trap-assisted tunneling (TAT), and band-to-band tunneling (BTBT).

We quantify the dark noise by calculating the 
DCR $R_D(T)$
via~\cite{xu2016}: 
\begin{equation}
R_D(T)= L \cdot \iint P_\textrm{trig}(\mathbf{r}) \cdot (G_\textrm{TAT}(\mathbf{r}, T) + G_\textrm{BTBT}(\mathbf{r}, T))~\dif\mathbf{r}
\label{eq:DCR}
\end{equation}
where $T$ is the temperature, $L$\,=\,16\,\si{\micro\meter} is the SPAD length, 
$P_\textrm{trig}(\mathbf{r})$ is the avalanche triggering probability, and 
$G_\textrm{TAT}(\mathbf{r}, T)$, $G_\textrm{BTBT}(\mathbf{r}, T)$ are the net generation rates of charge carriers (per unit volume) of their respective mechanisms. 

\subsubsection{Trap-Assisted Tunneling}
The thermal generation rate of carriers can be obtained from the Shockley-Read-Hall (SRH) model, modified to account for TAT~\cite{Kindt1998,Hurkx1992}:

\begin{equation}
G_{\textrm{TAT}}(\mathbf{r},T) = \frac{n_i(T)}{\tau_g(\mathbf{r},T)}
\label{eq:G_srh-tat}
\end{equation}
where $n_i(T)$ is the intrinsic carrier concentration and $\tau_g(\mathbf{r},T)$ is the electron-hole pair generation lifetime, which can be expressed in terms of the recombination lifetime $\tau_r(\mathbf{r},T)$~\cite{Schroder1997}:

\begin{equation}
\tau_g(\mathbf{r},T) = \frac{\tau_r(\mathbf{r},T) \cdot e^{\left|E_t - E_i\right|/k_BT}}{1+\Gamma(\mathbf{F}(\mathbf{r}),T)} 
\label{eq:tau_g}
\end{equation}
where the exponential term describes the main temperature dependence in TAT, and the field effect function $\Gamma(\mathbf{F}(\mathbf{r}),T)$ describes the effect of electric fields.
$E_t$ and $E_i$ are the energy levels of the recombination centers (assumed to be equal to that of traps at the Si/SiO$_2$ interface~\cite{Ling1997}) and the intrinsic Fermi level, respectively, and $k_B$ is the Boltzmann constant.

The field effect function $\Gamma(\mathbf{F}(\mathbf{r}),T)$ is:
\begin{equation}
\Gamma(\mathbf{F}(\mathbf{r}),T) = 2\sqrt{3\pi} \cdot \frac{|\mathbf{F}(\mathbf{r})|}{F_{\Gamma}(T)} \cdot \exp\left(\left(\frac{|\mathbf{F}(\mathbf{r})|}{F_{\Gamma}(T)}\right)^{2}\right)
\label{eq:Gamma}
\end{equation}
in which
\begin{equation}
F_{\Gamma}(T) = \frac{\sqrt{24 m_t^{*} (k_BT)^{3}}} {q \hbar}
\label{eq:F-gamma}
\end{equation}
where 
$q$ is the electron charge, and $m_t^{*}=0.25\,m_0$ is the effective electron tunneling mass, with $m_{0}$ being the electron rest mass~\cite{Hurkx1992-2}. 

\subsubsection{Band-to-Band Tunneling}
The BTBT mechanism has been shown to be important at electric field strengths above \mbox{$7 \times 10^5$~V$/$cm},
where band-bending is sufficiently strong to allow significant tunneling of electrons from the valence band to the conduction band~\cite{Hurkx1992-2}.
This rate can be expressed as:
\begin{equation}
G_{\textrm{BTBT}}(\mathbf{r},T) = B_A \cdot |\mathbf{F}(\mathbf{r})|^{B_{\Gamma}} \cdot \exp\Big(\frac{-B_B(T)}{|\mathbf{F}(\mathbf{r})|}\Big)
\label{eq:G_btbt}
\end{equation}
where $B_A$, $B_B$, and $B_{\Gamma}$ are model parameters; we use values based on ref~\cite{Hurkx1992-2}. 

The values of the parameters used in our calculations are listed in Table~\ref{table:all}, and further details of their derivation can be found in the Appendix.

\subsubsection{Avalanche Triggering Probability}
\begin{figure}[tb]
	\centering
	\includegraphics[width=1\linewidth]{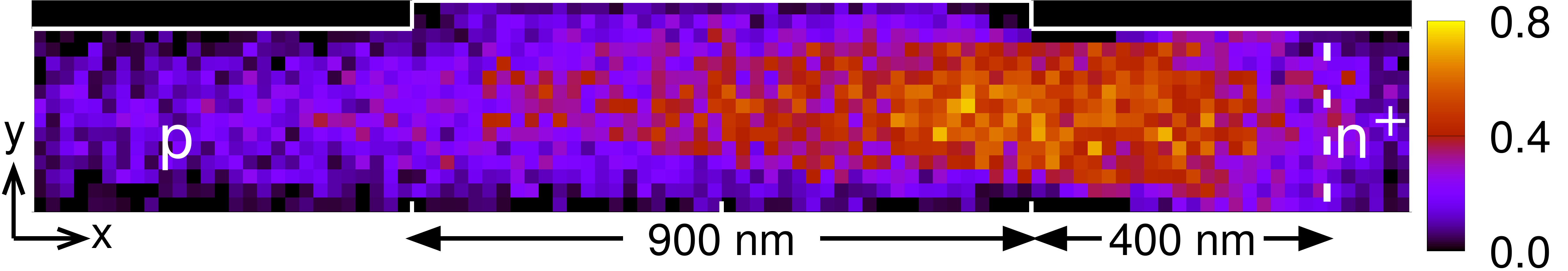}\\
	\caption{Avalanche triggering probability $P_\textrm{trig}(\mathbf{r})$ for a p-n$^{+}$ SPAD with $\Delta j$\,=\,400\,nm at $V_B$\,=\,16.5\,V, obtained over $>$\,40k Monte Carlo simulation runs. 
	Each 20$\times$20\,nm pixel shows the probability of an initial photo-generated electron-hole pair injected within that pixel resulting in a successful detection event. 
	The dashed line indicates the junction position.}
	\label{fig:Avalanche_triggering_probability}
\end{figure}

To obtain the avalanche triggering probability $P_\textrm{trig}(\mathbf{r})$ for each device, 
we perform $>$\,40k Monte Carlo simulation runs, with photon absorption positions distributed uniformly across the device.
A representative map of $P_\textrm{trig}(\mathbf{r})$ is shown in Fig.~\ref{fig:Avalanche_triggering_probability}.

\section{Simulator Optimization}
\label{sec:simulation_optimization}

The Monte Carlo simulations can become computationally expensive due to the need to keep track of and model individual charge carriers, 
especially when the number of charge carriers grows exponentially during the avalanche process.
If we would use the same simulation parameters 
in our previous work~\cite{Yanikgonul2018} to model one SPAD at a given bias $V_B$, our simulator (implemented in Python) would require $\sim$24k CPU-hours on two sets of 12-core CPUs (Intel\textsuperscript \textregistered ~Xeon\textsuperscript \textregistered~\mbox{E5-2690 v3}).
Such a high computation cost would limit the variety of SPAD designs we can feasibly study.

Thus, we first use a representative device (p-n$^{+}$ SPAD with $\Delta j$\,=\,-50\,nm, at $V_B$\,=\,21.5\,V) 
to perform a series of preliminary studies to optimize the simulation parameters: the detection threshold~$I_\textrm{det}$, RPL time step~$\Delta t_{\textrm{rpl}}$, and number of simulation runs per parameter set.
We aim to reduce computation time without sacrificing the simulation accuracy.

\begin{figure}[tb]
    \centering
    \includegraphics[width=0.90\linewidth]{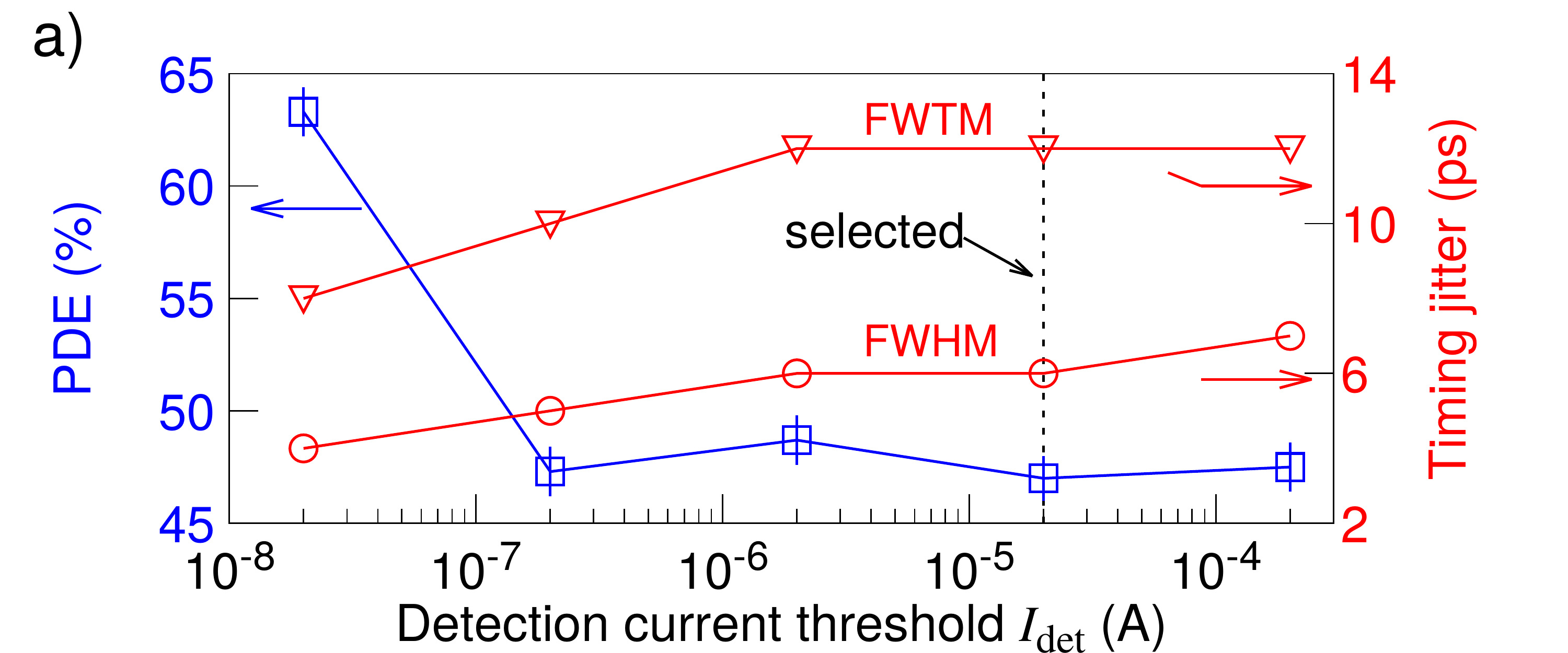}\\
    \includegraphics[width=0.90\linewidth]{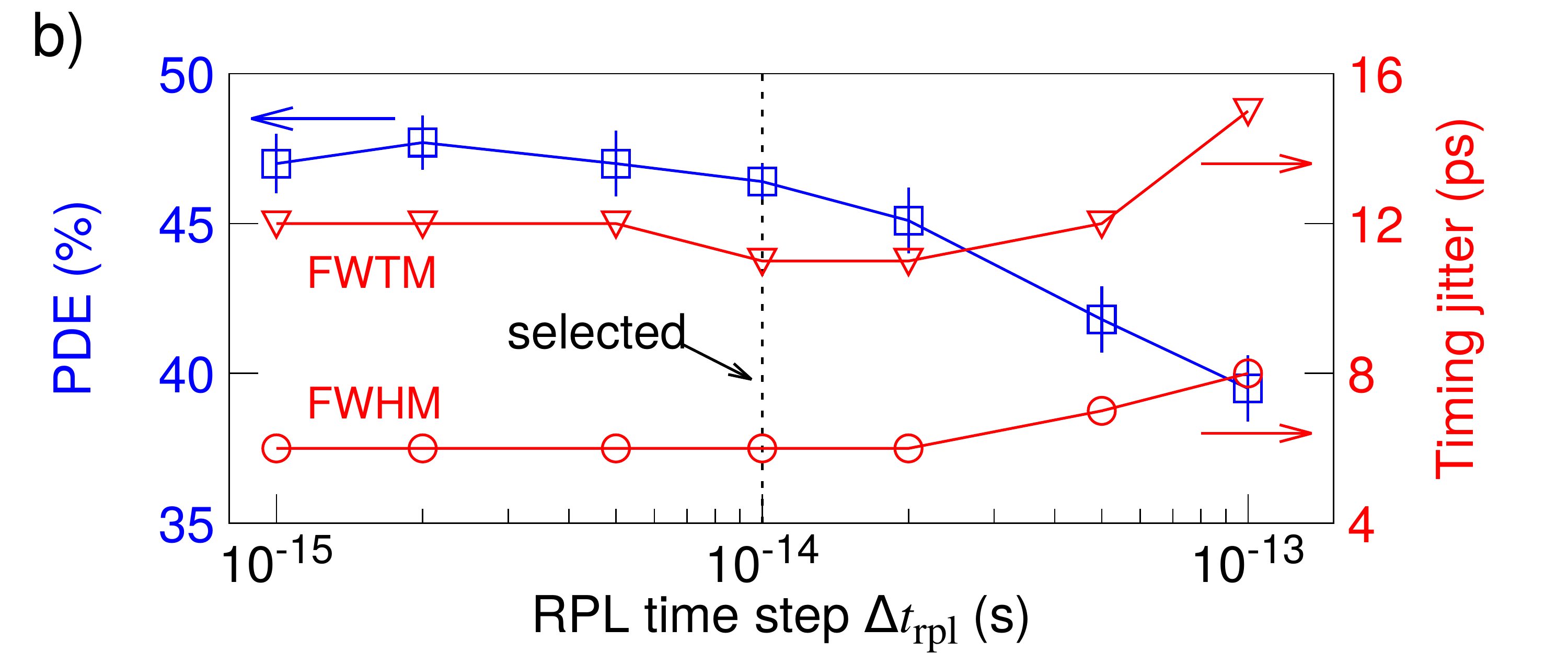}\\
    \includegraphics[width=0.90\linewidth]{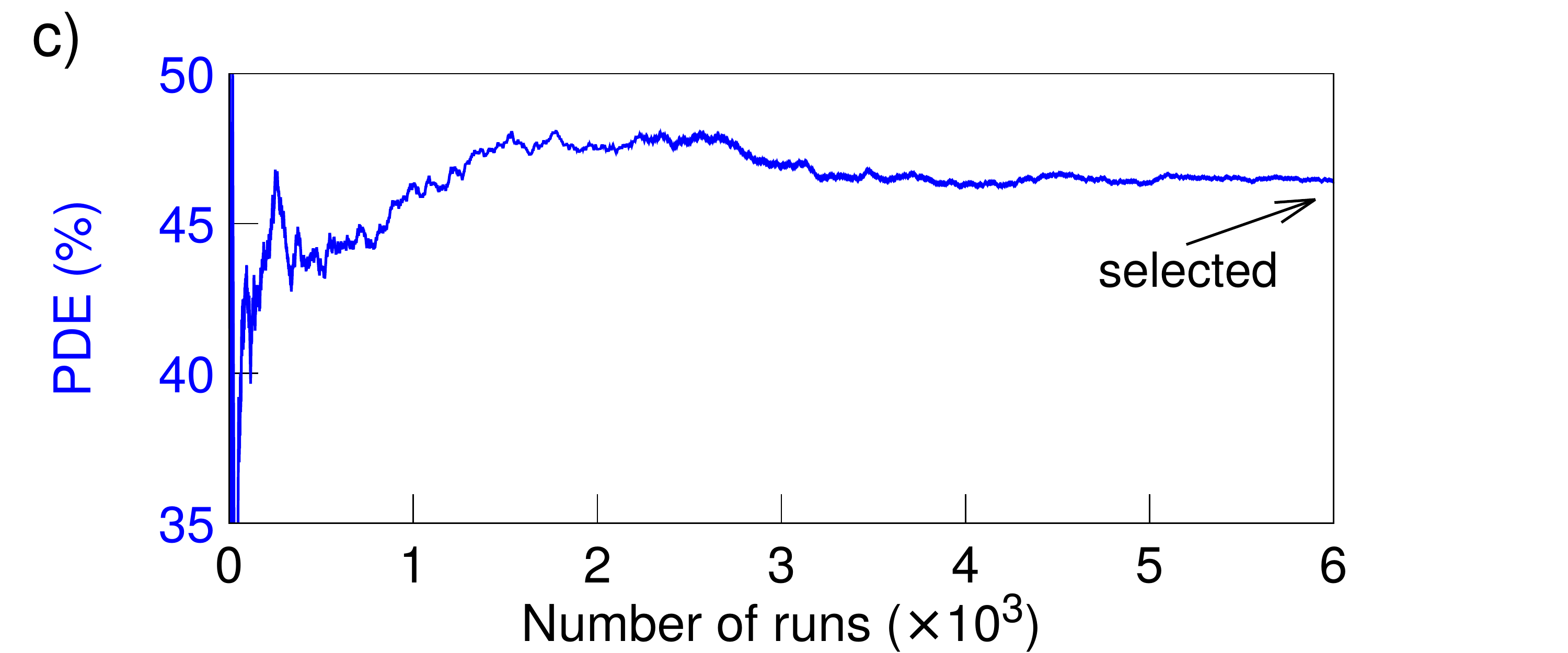}
    \caption{Optimization study of varying simulator parameters and their effects on the photon detection efficiency (PDE) and timing jitter (full-width-half-max (FWHM) and full-width-tenth-max (FWTM)), for a p-n$^{+}$ SPAD with \mbox{$\Delta j$\,=\,-50\,nm} at $V_B$\,=\,21.5\,V. 
    (a)~Varying $I_\textrm{det}$ with $\Delta t_{\textrm{rpl}}$\,=\,1\,fs, 2k simulation runs per $I_\textrm{det}$ value.
    (b)~Varying $\Delta t_{\textrm{rpl}}$ with $I_\textrm{det}$\,=\,20\,\si{\micro\ampere}, 2k simulation runs per $\Delta t_{\textrm{rpl}}$ value. 
    (c)~Convergence of PDE for $\Delta t_{\textrm{rpl}}$\,=\,10\,fs and $I_\textrm{det}$\,=\,20\,\si{\micro\ampere} after several thousand runs.
    Error bars for PDE indicate 1\,s.d. uncertainty.
    Selected parameters for subsequent simulations in this paper are marked. 
    }
    \label{fig:Simulation_optimization}
\end{figure}

\subsection{Detection Current Threshold}

A reasonable discriminator threshold in experimental SPAD characterization setups is $I_\textrm{det}$\,=\,0.2\,mA~\cite{Spinelli1997},
a value we used previously~\cite{Yanikgonul2018}. 
However, it may not be necessary to simulate the multiplication of charge carriers up to that point
as the avalanche process might already have passed a self-sustaining threshold at a lower current. 
On the other hand, a very low~$I_\textrm{det}$ would overestimate the PDE by falsely identifying small avalanches that would not be self-sustaining, 
and underestimate the timing jitter by not simulating the full avalanche.

By varying~$I_\textrm{det}$ while fixing $\Delta t_{\textrm{rpl}}$\,=\,1\,fs 
with 2k simulation runs per $I_\textrm{det}$ value (Fig.~\ref{fig:Simulation_optimization}(a)),
we conclude that we can lower~$I_\textrm{det}$ to 20\,\si{\micro\ampere} without significant deviations in PDE or timing jitter.

\subsection{RPL Time Step}

A larger RPL time step $\Delta t_{\textrm{rpl}}$ would speed up simulations, but reduces time resolution and hence accuracy.
A suitable choice 
would be just short enough such that the charge carrier environment does not change too significantly between each step, even in the high-field regions with large field gradients.

We vary $\Delta t_{\textrm{rpl}}$ while fixing $I_\textrm{det}$\,=\,20\,\si{\micro\ampere} 
with 2k simulation runs per $\Delta t_{\textrm{rpl}}$ value (Fig.~\ref{fig:Simulation_optimization}(b)).
We choose $\Delta t_{\textrm{rpl}}$\,=\,10\,fs as an optimal value; for larger time steps, PDE begins to deviate significantly compared to the previous value of $\Delta t_{\textrm{rpl}}$\,=\,1\,fs.

\subsection{Number of Simulation Runs}

We analyze the PDE over an increasing number of simulation runs for $\Delta t_{\textrm{rpl}}$\,=\,10\,fs and $I_\textrm{det}$\,=\,20\,\si{\micro\ampere}, and observe that the PDE converges to a stable value after several thousand runs.
We choose to perform at least 6k runs per parameter set to reduce the relative uncertainty to~$\sim$1\%.

Compared to the previous simulation parameters (i.e. $\Delta t_{\textrm{rpl}}$\,=\,1\,fs, $I_\textrm{det}$\,=\,0.2\,mA, 18k runs), 
our optimized values ($\Delta t_{\textrm{rpl}}$\,=\,10\,fs, $I_\textrm{det}$\,=\,20\,\si{\micro\ampere}, 6k runs) require only $\sim$\,90 CPU-hours per set,
indicating an improved timing performance by a factor of $\sim$\,270.

\begin{table}[tb]
\centering
\caption{Simulation Parameters}
\label{table:all}
\begin{tabular}{l|l|l|l}
\hline
\textbf{Name} & \textbf{Symbol} & \textbf{Value} & \textbf{Reference} \\
\hline
Electric field threshold & $F_{\textrm{thr}}$  & 1 $\times$ 10$^\textrm{5}$ V cm$^\textrm{-1}$ & \cite{Wegrzecka}\\
Detection current threshold & $I_{\textrm{det}}$  & 20\,\si{\micro\ampere} & - \\
RPL time step            & $\Delta t_{\textrm{rpl}}$ & 10\,fs            & - \\
No. of simulations per   &          -          &   $>$\,6000      & - \\
parameter set        &                     &                        &   \\
BTBT parameter           & $B_A$   & 4$\times$ 10$^\textrm{14}$     & \cite{Hurkx1992-2} \\
                         &                     &cm$^\textrm{-0.5}$V$^\textrm{-2.5}$s$^\textrm{-1}$ &  \\ 
                         & $B_{\Gamma}$       & 2.5                    & \cite{Hurkx1992-2} \\
Recombination energy & $E_t$-$E_i$ & 0.25 eV & \cite{Ling1997} \\
&&&\\
Temperature dependent    &                     &                        &                   \\
parameters at 300 (243)\,K:&                    &                        &                   \\
-- intrinsic carrier     & $n_i$    &  9.70$\times$10$^\textrm{9}$            & \cite{Misiakos1993}\\
~~~concentration         &                     & (2.95$\times$10$^\textrm{7}$) cm$^\textrm{-3}$      & \\
-- Recombination lifetime & $\tau_r$ & 7.0 (7.8) ns        & \cite{Park2010,Ling1997}\\  
-- BTBT parameter       & $B_B$      & 1.90 (1.94) $\times$ 10$^\textrm{7}$   & \cite{Hurkx1992-2,Bludau1974} \\
                         &                     & V cm$^\textrm{-1}$             &  \\

\hline
\end{tabular}
\end{table}

\section{Simulation Results and Discussion}
\subsection{Photon Detection Efficiency and Timing Jitter}

\begin{figure}[tb]
    \centering
    \includegraphics[width=0.75\linewidth]{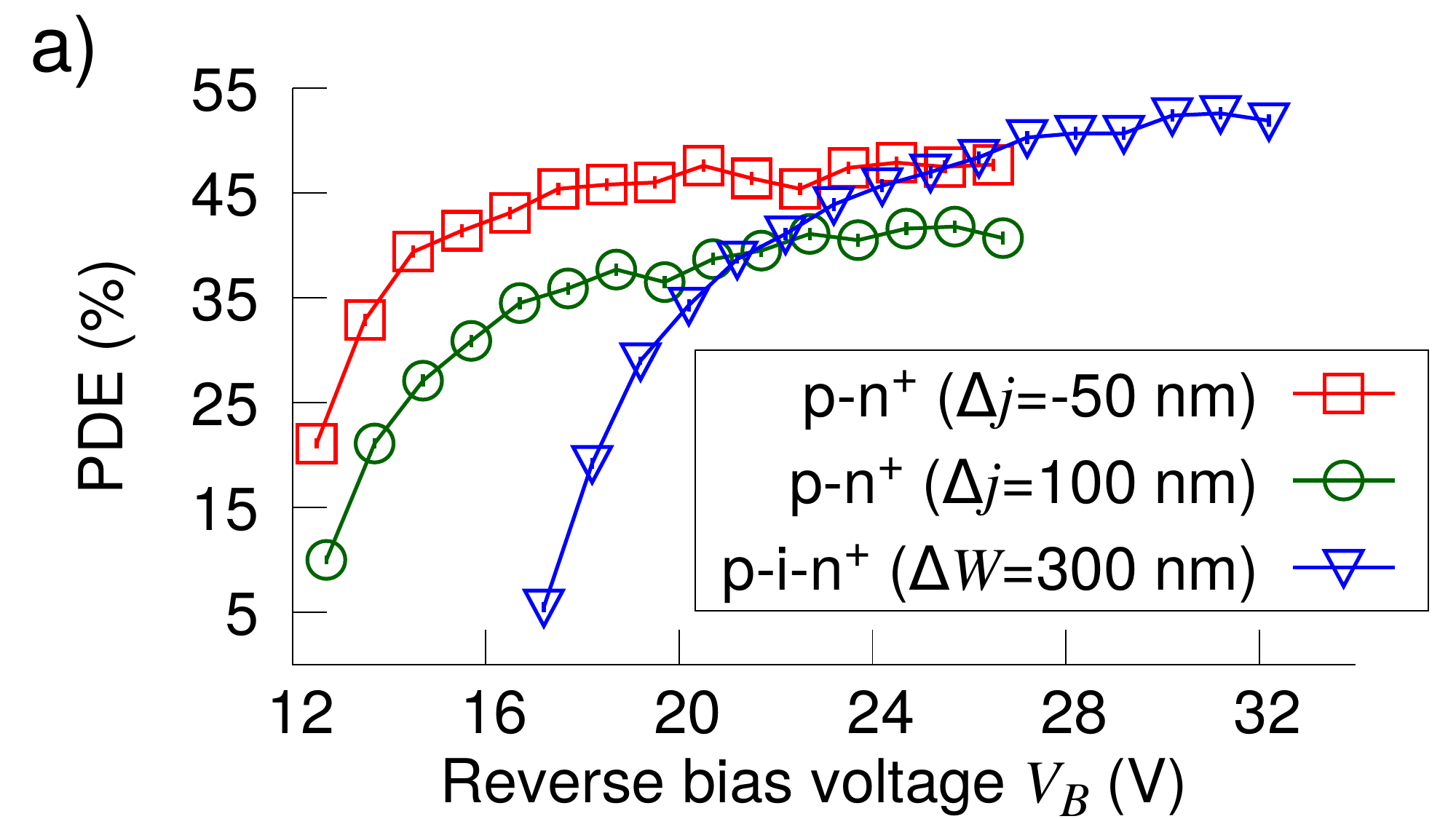}\vspace{0.2cm}
    \includegraphics[width=0.49\linewidth]{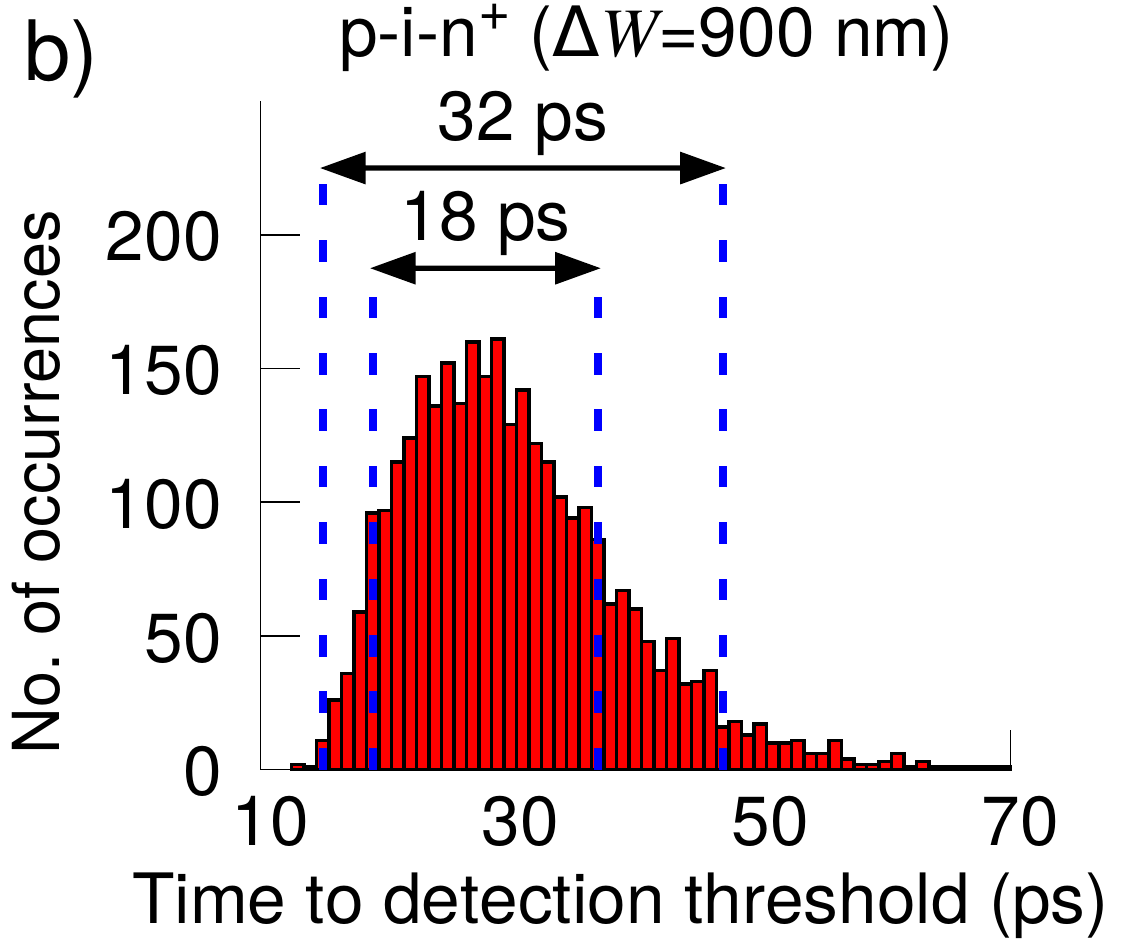}
    \includegraphics[width=0.49\linewidth]{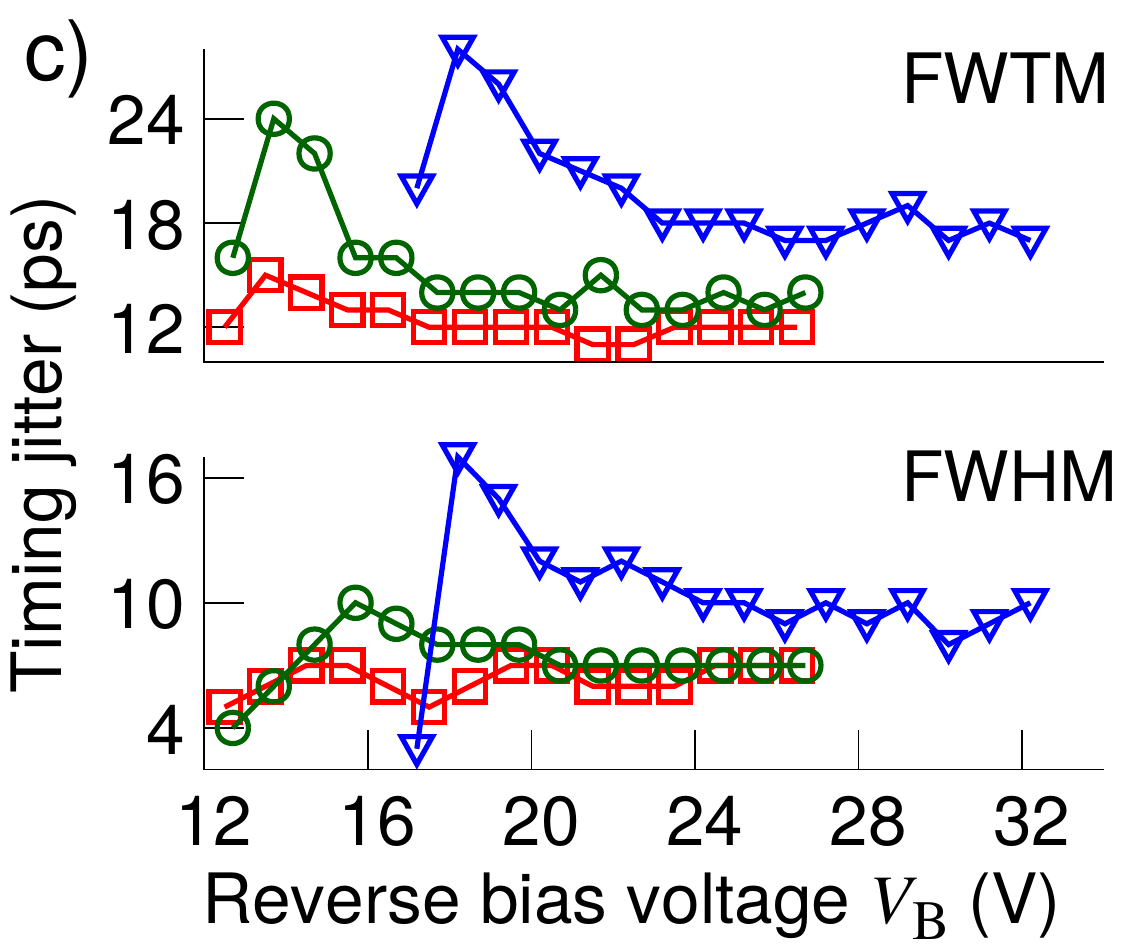}
    \caption{(a) Simulated PDE at varying reverse bias voltages $V_B$ for representative devices, showing the saturation behavior as $V_B$ increases.
    Error bars indicating 1\,s.d. uncertainty are much smaller than the symbol size.
    (b) Distribution of simulated avalanche times (i.e. time between photon absorption and reaching the detection threshold $I_\textrm{det}$)
    for a \mbox{p-i-n$^{+}$} SPAD with \mbox{$\Delta W$\,=\,900\,nm} at $V_B$\,=\,41\,V. Histogram bin size is 1\,ps. The full-width-half-max (FWHM) and full-width-tenth-max (FWTM) timing jitter values are indicated. 
    (c) FWHM and FWTM timing jitter performance for the same devices in (a). (a) and (c) share the same legend.}
    \label{fig:PDE_vs_bias}
\end{figure}

We simulate each device at increasing reverse bias voltages~$V_B$, starting from just above its breakdown voltage.
For all devices, PDE increases with $V_B$ and reaches a saturation level (representative plots shown in Fig.~\ref{fig:PDE_vs_bias}(a)).
We define the saturated bias voltage $V_\textrm{sat}$ as the lowest $V_B$ value where the obtained PDE values within a $\pm$1\,V range agree within their 1\,s.d. uncertainty;
the PDE at $V_\textrm{sat}$ is then the saturated PDE.

The distribution of avalanche times is generally asymmetric, 
especially for \mbox{p-i-n$^+$} SPADs with $\Delta W$\,$>$\,600\,nm (see Fig.~\ref{fig:PDE_vs_bias}(b)).
Long tails in the timing distribution can adversely affect applications requiring high timing accuracies, e.g. satellite-based quantum communications~\cite{Agnesi:19}.
Therefore, we present the full-width-half-maximum (FWHM) and full-width-tenth-maximum (FWTM) timing jitter,
both extracted from timing histograms with 1\,ps bin size,
to better describe the timing performance of the SPADs.
In general, timing jitter does not vary significantly with $V_B$, except when $V_B$ is near the breakdown voltage.

\begin{figure}[tb]
    \centering
    \includegraphics[width=0.9\linewidth]{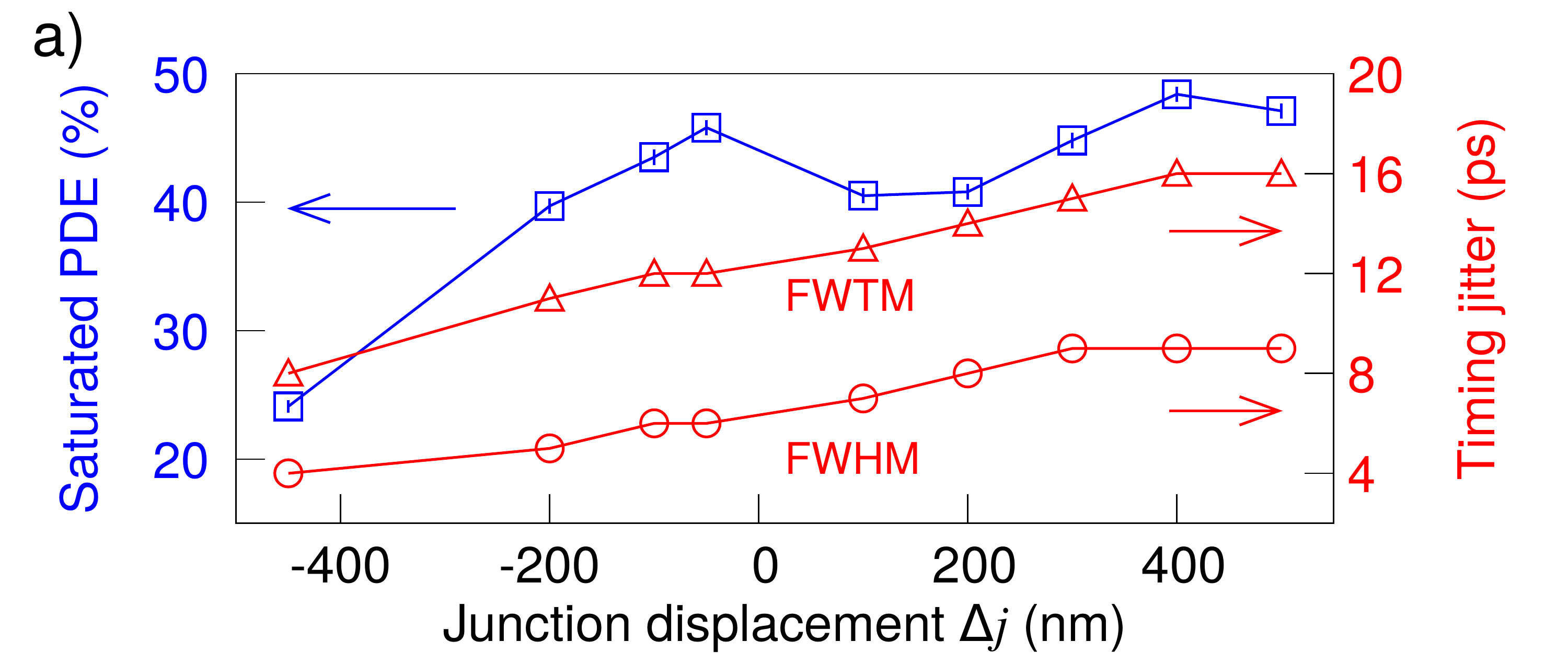}\\
    \includegraphics[width=0.9\linewidth]{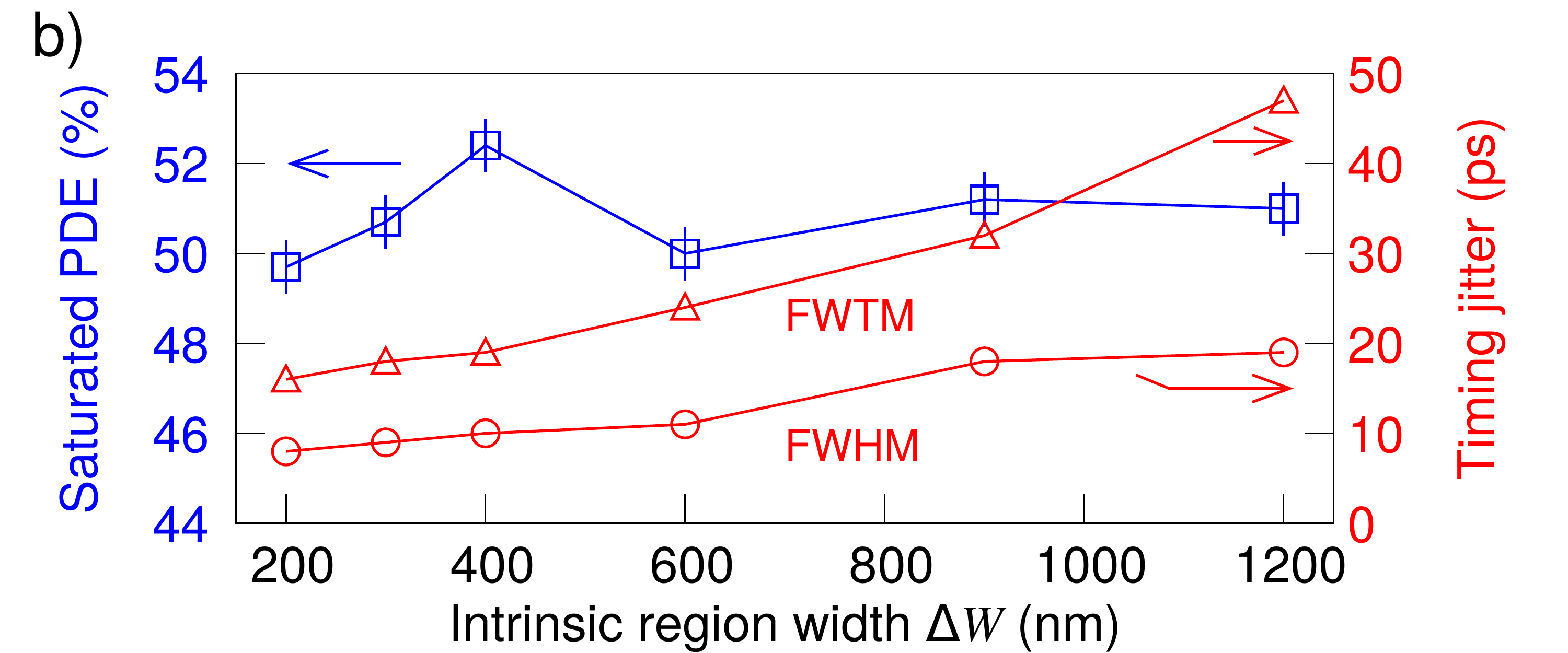}\\
    \caption{Saturated PDE and timing jitter for various (a) p-n$^{+}$ and (b) p-i-n$^{+}$ SPADs.
    Both full-width-half-max (FWHM) and full-width-tenth-max (FWTM) timing jitter values are shown.
    Error bars for PDE indicate 1\,s.d. uncertainty.}
    \label{fig:PDE_vs_design_parameter}
\end{figure}

\subsubsection{p-n$^{+}$ SPADs}
For p-n$^{+}$ SPADs, we observe a general trend of PDE increasing with the junction displacement~$\Delta j$ (Fig.~\ref{fig:PDE_vs_design_parameter}(a)). 
If the junction is placed further away from the waveguide core, charge carriers injected after a photon absorption in the core region travel a longer distance and can undergo more impact ionizations, thus increasing the likelihood of a successful avalanche.
The stochastic avalanche process taking place over a larger distance would also explain the increasing timing jitter at higher~$\Delta j$.
However,~$\Delta j$ being too large would weaken the electric field strength in the waveguide core, which would lead to more charge carriers being lost at the waveguide boundaries due to random walk; 
this may explain the slight drop in PDE for~$\Delta j>$~400\,nm. 

The observed drop in PDE for~$\Delta j$\,=\,100\,nm is due to an ``edge effect'': 
when the junction is placed in close proximity to the waveguide rib edge,
we observe a narrowing of the effective impact ionization region where ionization coefficients are high~(Fig.~\ref{fig:Electric_fields}(b)), 
which leads to a lower PDE.

The highest saturated PDE obtained for p-n$^{+}$ SPADs is 48.4\,$\pm$\,0.6\% at $V_B$\,=\,26.5\,V for $\Delta j=400$ nm, with a FWHM timing jitter of 9\,ps.

\subsubsection{p-i-n$^{+}$ SPADs}
For p-i-n$^{+}$ SPADs, 
the widening of the high-field region has led to a higher PDE than for \mbox{p-n$^{+}$} devices (Fig.~\ref{fig:PDE_vs_design_parameter}(b)). 
Besides the increased efficiency of impact ionizations, this can also be explained by a lower loss rate of charge carriers under the diffusion model ($<$5\% for \mbox{p-i-n$^{+}$}, $\sim$10\% for \mbox{p-n$^{+}$}), which follows a photon absorption event in the quasi-neutral regions.  
We do not find an obvious dependence of the PDE on the intrinsic region width for~$\Delta W>$~400\,nm, 
although timing jitter increases with~$\Delta W$.

Based on our analysis, we conclude that the optimum performance is obtained for~$\Delta W$\,=\,400\,nm, which gives a saturated PDE of 52.4\,$\pm$\,0.6\% at $V_B$\,=\,31.5\,V and a FWHM timing jitter of 10\,ps.

\subsection{Dark Count Rate}

\begin{figure}[tb]
    \centering
    \includegraphics[width=0.9\linewidth]{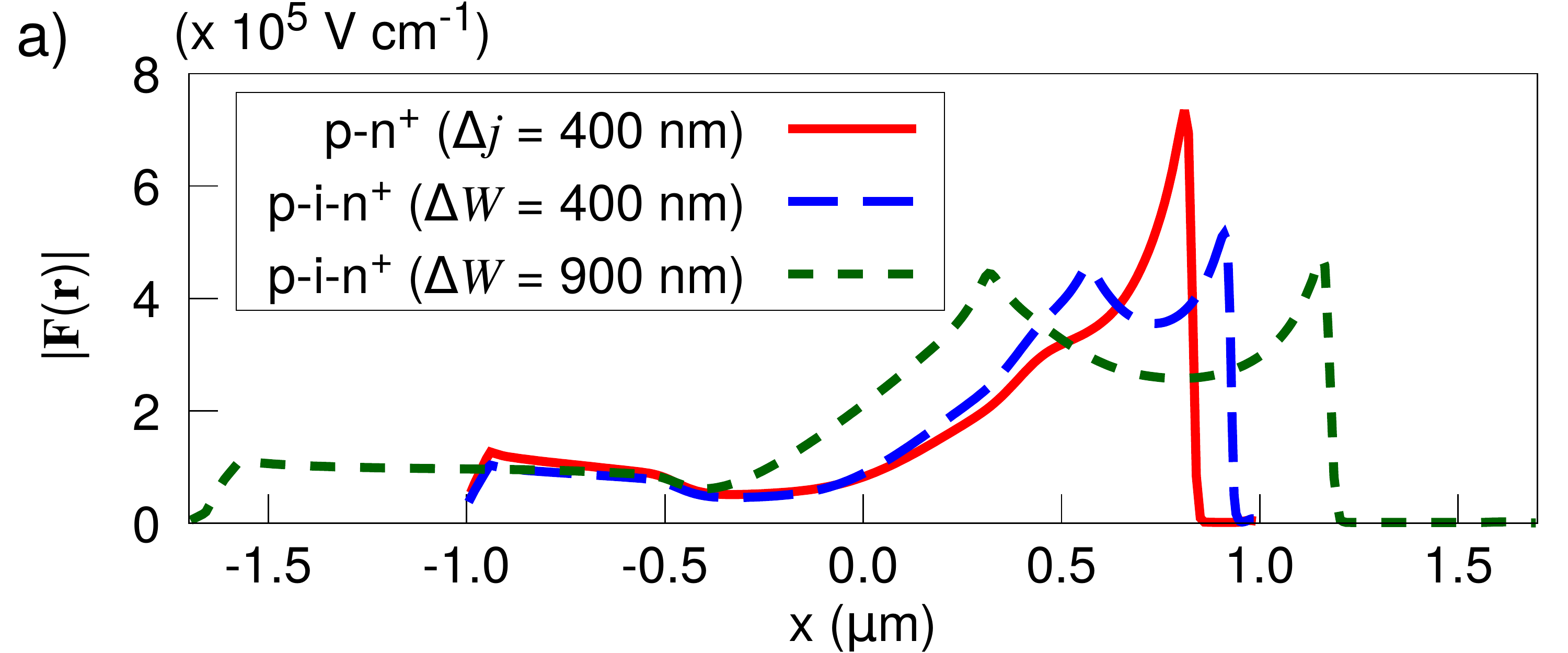}\vspace{-0.2cm}
    \includegraphics[width=0.9\linewidth]{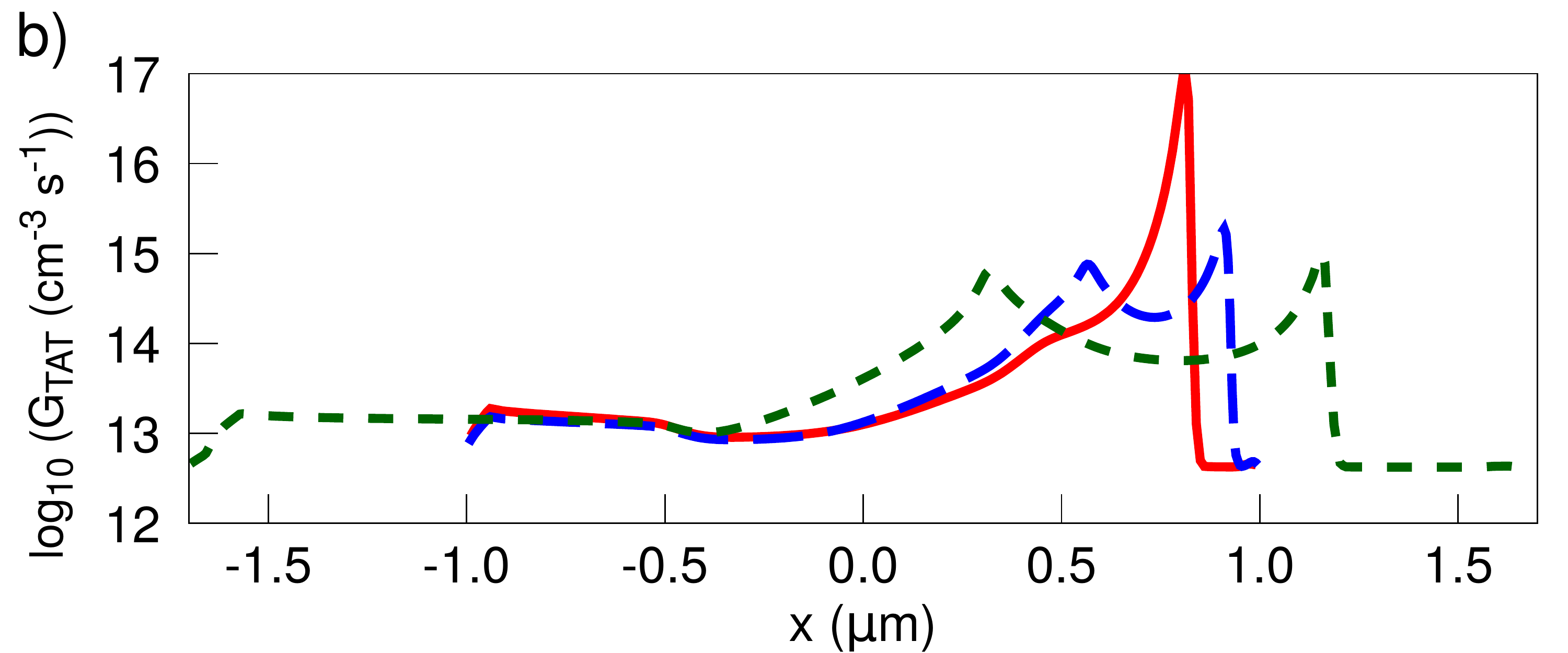}\vspace{-0.2cm}
    \includegraphics[width=0.9\linewidth]{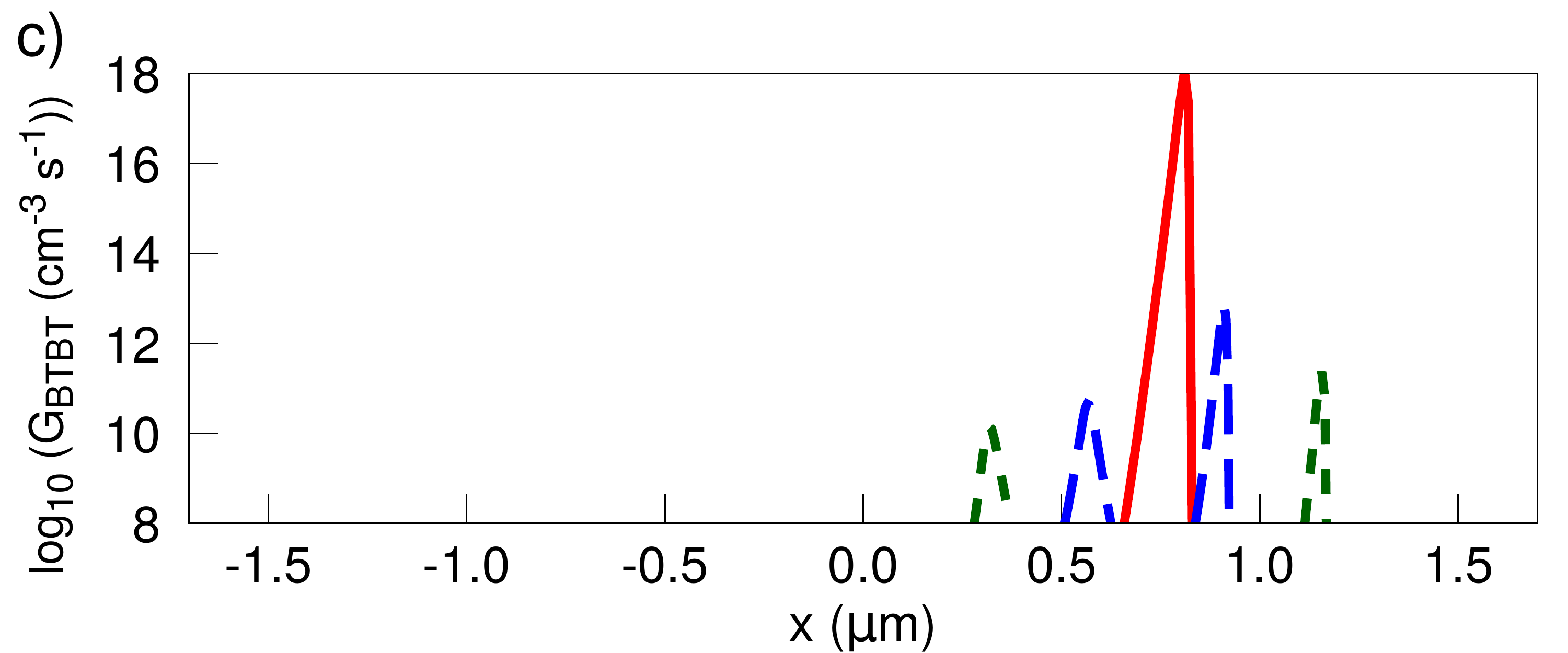}
    \caption{(a) Electric field strength $|\mathbf{F}(\mathbf{r})|$, (b) trap-assisted tunneling (TAT) generation rates, and (c) band-to-band tunneling (BTBT) generation rates at the waveguide mid-height (170\,nm from the bottom) for SPADs with high saturated PDEs:
    p-n$^{+}$ SPAD with $\Delta j$\,=\,400\,nm,
    and p-i-n$^{+}$ SPADs with $\Delta W$\,=\,400\,nm and 900\,nm,
    at reverse bias voltages beyond where their PDE has already saturated ($V_B$\,=\,31.5\,V, 34.5\,V, and 53\,V, respectively).
     } 
    \label{fig:isoline_data}
\end{figure}

We also evaluate the dark noise performance of the SPADs, focusing on devices which display high saturated PDE: 
\mbox{p-n$^{+}$} SPAD with $\Delta j$\,=\,400\,nm and \mbox{p-i-n$^{+}$} SPADs with $\Delta W$\,=\,400\,nm and 900\,nm.
We calculate the DCR at 243\,K, which is in a typical SPAD operating regime readily achieved with thermoelectric cooling, as well as at 300\,K to explore the feasibility of room temperature operation.

\begin{figure}[tb]
    \centering
    \includegraphics[width=1\linewidth]{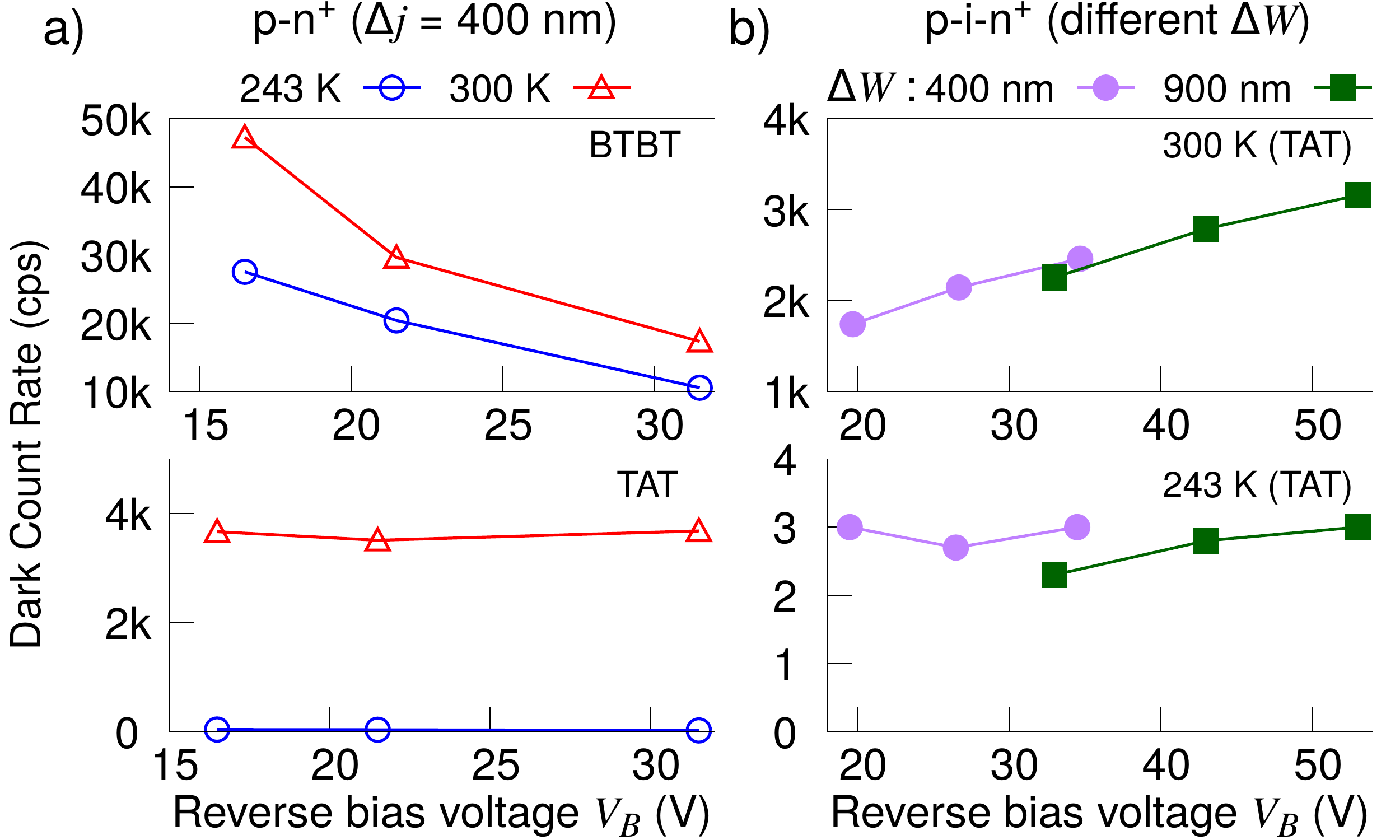}\\
    \caption{Dark count rate (DCR) contributions due to TAT and BTBT mechanisms at varying reverse bias $V_B$ and temperatures. 
    (a) \mbox{p-n$^{+}$} SPAD with $\Delta j$\,=\,400\,nm. 
    (b) \mbox{p-i-n$^{+}$} SPADs with $\Delta W$\,=\,400\,nm and 900\,nm. The contribution from BTBT is negligible, thus only TAT is shown here.
    }
    \label{fig:DCR_at_T}
\end{figure}

For our simulated parameters, BTBT shows a greater sensitivity to peak electric field strength than TAT~(Fig.~\ref{fig:isoline_data}).
In \mbox{p-n$^{+}$} SPADs, where the peak fields are high, BTBT is the dominant dark carrier generation mechanism. 
As the bias $V_B$ increases, the depletion region widens, leading to a decrease in the peak field strength and hence the overall DCR, while the TAT contribution stays relatively constant~(Fig.~\ref{fig:DCR_at_T}(a)). 
At an operating bias of $V_B$\,=\,31.5\,V (which is above the saturated bias), the DCR is 11\,kcps and 21\,kcps at 243\,K and 300\,K, respectively.

In \mbox{p-i-n$^{+}$} SPADs, due to wider high-field regions with lower peak fields, BTBT becomes negligible compared to TAT.
As such, DCR generally increases with $V_B$, and shows a steeper dependence on temperature ($\sim$\,1000\,-\,fold drop between 300\,K and 243\,K).
We observe that while SPADs with wider intrinsic region widths $\Delta W$ had lower dark carrier generation rates per unit volume, this was offset by the larger device volume, and could lead to higher DCR compared to narrower $\Delta W$.

Overall, dark count performance for \mbox{p-i-n$^{+}$} SPADs is significantly better compared to \mbox{p-n$^{+}$} devices, 
with observed DCR of $<$\,4\,kcps at 300\,K and $<$\,5\,cps at 243\,K~(Fig.~\ref{fig:DCR_at_T}(b)), even at $V_B$ beyond the saturated bias.

\section{Conclusions}

In conclusion, we have simulated waveguide-based silicon SPADs for visible wavelengths, 
studying both \mbox{p-n$^{+}$} and \mbox{p-i-n$^{+}$} doping profiles.
For our simulated parameters, \mbox{p-i-n$^{+}$} SPADs outperform \mbox{p-n$^{+}$} devices in terms of PDE and DCR;
we identify the optimum device as a \mbox{p-i-n$^{+}$} SPAD with $\Delta W$\,=\,400\,nm, 
with a saturated PDE of 52.4\,$\pm$\,0.6\% at a bias of $V_B$\,=\,31.5\,V, FWHM timing jitter of 10\,ps, and DCR $<$\,5\,cps at 243\,K.
This is an improvement over our previous study, where the highest PDE obtained was~45\%~\cite{Yanikgonul2018}.

The PDE is slightly lower than typical free-space SPAD modules with PDEs of up to $\sim$\,70\%~\cite{eisaman2011invited};
however, our waveguide devices can offer superior timing performance and dark noise compared to available commercial devices (jitter $\sim$\,35\,ps, DCR $<$\,25\,cps). 
We note that even at room temperature, the DCR of a few kcps is acceptable for certain important technologies including LIDAR~\cite{takai2016single} due to the use of temporal gating,
thus indicating the potential applicability of our waveguide SPADs.

Our simulation methods can also be further extended to study other device geometries (e.g. trapezoid waveguides), doping profiles (e.g. p$^+$-i-p-n$^+$) and materials (e.g. Ge-on-Si SPADs for near-infrared wavelengths).

\appendix
\label{sec:Appendix}
\subsection{Trap-Assisted Tunneling}

\subsubsection{Intrinsic carrier concentration}

We calculate the intrinsic carrier concentration $n_i(T)$ in silicon via~\cite{Misiakos1993}:
\begin{equation}
n_i(T)=5.29\times10^{19} \cdot (T/300)^{2.54} \cdot \exp(-6726/T)
\label{eq:Intrinsic_carrier_concentration}
\end{equation}

\subsubsection{Effective Recombination Lifetime}

The effective recombination lifetime 
$\tau_r(T)$ was measured to be 7\,ns at room temperature
for an undoped Si rib waveguide device with similar sub-\si{\micro\metre} dimensions~\cite{Park2010}.
To obtain a suitable value at 243\,K, we analyze the temperature dependence of $\tau_r(T)$:
for low-level injection in p-type silicon, $\tau_r(T)$ can be approximated as the electron recombination lifetime~\cite{Schroder1997}, i.e.:
\begin{equation}
	\tau_r(T) \approx \frac{1}{\sigma_e \cdot \nu_e(T) \cdot N_t}
\label{eq:tau_n}
\end{equation}
where $\sigma_e$ is the electron capture cross section, $\nu_e(T)$ is the mean thermal velocity of electrons, and $N_t$ is the trap density.
The trap density $N_t$ is assumed to be temperature-independent, while for traps at Si/SiO$_{2}$ interface with \mbox{$E_t$\,-\,$E_i$\,=\,0.25\,eV}, $\sigma_e$~has been shown to be relatively constant over our relevant temperature range (243\,--\,300\,K)~\cite{Garetto2012}.
Thus, the temperature dependence comes only from $\nu_e(T) \propto \sqrt{T}$, and we obtain
\begin{equation}        
\tau_r(243) = \tau_r(300) \cdot \sqrt{300/243}
\label{eq:tau_n_lower_temp}
\end{equation}

\subsection{Band-to-Band Tunneling}
Values for $B_A$, $B_B$ and $B_{\Gamma}$ at room temperature are given in ref.~\cite{Hurkx1992-2}.
Both $B_A$ and $B_{\Gamma}$ are nominally temperature-insensitive, 
while $B_B(T) \propto [E_g(T)]^{3/2}$, where $E_g(T)$ is the Si bandgap energy~\cite{Bludau1974}:
\begin{equation}
E_g(T) = A + BT + CT^2
\label{eq:bandgap}
\end{equation}
in which $A=1.1785$ eV, $B=-9.025 \times 10^{-5}$ eV/K, and \mbox{$C=-3.05 \times 10^{-7}$ eV/K$^2$}, for 150~K $\leq T \leq$ 300~K.
We thus obtain:
\begin{equation}
B_B(243) = B_B(300) \cdot \Big(\frac{E_g(243)}{E_g(300)}\Big)^{(3/2)}
\label{eq:B_B}
\end{equation}

\section*{Acknowledgment}

The authors acknowledge the usage of computational resources of the National Supercomputing Centre, Singapore (https://www.nscc.sg) for this work.

\bibliographystyle{IEEEtran}
\bibliography{IEEEabrv,References/SPAD_simulation_references}

\end{document}